\documentclass[journal=jpcbfk,manuscript=article]{achemso}
\usepackage{longtable}
\usepackage{multirow}
\usepackage{amsmath}
\usepackage[usenames,dvipsnames]{color}
\usepackage{soul}
\usepackage{threeparttable}
\usepackage{xr}
\usepackage{graphicx}
\usepackage{amsmath,amssymb}
\usepackage{caption}
\usepackage{color}
\usepackage[table]{xcolor}
\usepackage{dcolumn}
\usepackage{bm}
\usepackage{float}
\usepackage{subcaption}
\usepackage{textcomp} \usepackage{url} \usepackage[normalem]{ulem}
\usepackage{comment}

 \usepackage{microtype}
\usepackage{etoolbox}

\usepackage[unicode=true, bookmarks=true, bookmarksnumbered=true,
  bookmarksopen=true, bookmarksopenlevel=2, breaklinks=false,
  pdfborder={0 0 1}, backref=false, colorlinks=true, hidelinks
]{hyperref}

\SectionNumbersOn

\author{Cangtao Yin} \author{Markus Meuwly} \affiliation[University of
  Basel]{Department of Chemistry, University of Basel,
  Klingelbergstrasse 80, CH-4056 Basel, Switzerland.}
\email{m.meuwly@unibas.ch}

\title{Reaction Pathway Detection using Machine-Learned Energy
  Potentials - Decomposition of Energized CF$_3$CHOO}

\begin{document}
\date{\today}

\begin{abstract}
Characterization of the decomposition products of energized Criegee
intermediates is essential for assessing their impact on the chemical
evolution of the atmosphere. Here, a generic and microscopically
resolved approach is used to determine the molecular fragmentation pathways and
products for CF$_3$CHOO. They
include, among others, direct formation of CO$_2$ + CHF$_3$ (HFC-23), HF + CO$_2$ +
CF$_2$, and fragmentation routes that are not evident from static
reaction path calculations alone. The computed probability for formation of HFC-23 of 14 \% qualitatively agrees with a value of $(7.9^{+0.4}_{-0.2})$ \% from recent measurements, given the differences in the two approaches. Non-statistical dynamics is found for almost all decomposition pathways and the simulations show that excess energy can redirect reaction outcomes away from minimum-energy pathways. The
results highlight the power of machine-learned PESs to elucidate
multi-step reaction mechanisms of atmospherically relevant
intermediates beyond traditional Master equation/electronic structure approaches to
provide molecular-level understanding of the role of dynamics.
\end{abstract}

\section{Introduction}
Criegee intermediates (CIs) of various
chemical composition play central roles in atmospheric
chemistry.\cite{criegee1949ozonisierung,marston:2008,khan:2018} These
compounds with the general chemical formula R$_1$R$_2$COO are relevant
due to both, their unimolecular decomposition products and the
bimolecular reactions they are involved in the atmosphere. The
species participate in a wide range of reactions that influence the
oxidative capacity of the atmosphere, aerosol formation, and the
budgets of trace
gases.\cite{chao2015direct,yin2017does,yin2018effect,yin2023nh3,yin2024revealing,cox2020evaluated}
In recent years, considerable effort has been devoted to understanding
the chemistry of substituted CIs, motivated by their relevance to both
atmospheric processes and industrial
emissions.\cite{cox2020evaluated,caravan2021open}\\

\noindent
A key characteristic of CIs is the notion that under atmospheric
conditions they are "born" with large amounts of internal energy
$(\sim 50\ {\rm kcal/mol})\cite{taatjes:2015}$ following alkene
ozonolysis, which differs from laboratory studies that generate CIs
e.g. through photolysis of diiodoalkanes in the presence of
O$_2$.\cite{hassan:2021} Traditionally, the R$_1$ and R$_2$ groups of
CIs include -H and -CH$_3$. However, over the past 10 or so
years,\cite{watson:2023} halogenated species involving FCHOO, ClCHOO,
CF$_3$CHOO, and CF$_3$CFOO have attracted increased attention. In particular, ozonolysis of hydrofluoro-olefins (HFOs) proceeds via analogous fluorinated Criegee intermediates, which has helped motivate growing interest in these halogenated systems. One of
the main reasons behind this development is the observation that
ozonolysis of HFOs also leads to long-lived
reaction products with high global warming potential
(GWP).\cite{mcgillen:2023} This finding challenges the
concept that HFOs are more environmentally benign alternatives to
hydrofluoro-carbons or related
compounds.\cite{mcgillen:2023,orr2025atmospheric}\\

\noindent
Chamber experiments have demonstrated that the ozonolysis of several
commercially relevant HFOs produces CHF$_3$ (HFC-23), a long-lived
species with a very high global warming potential, with yields that
can reach several percent depending on molecular structure and
reaction conditions.\cite{mcgillen:2023,orr2025atmospheric}
Similar behavior has been observed for related
hydrochlorofluoroolefins (HCFOs), for which measurable CHF$_3$
formation occurs during ozonolysis. This highlights the broader
relevance of this pathway across fluorinated olefin
classes.\cite{nielsen2025cf3h} In addition to CHF$_3$, other
persistent products including CF$_4$ (PFC-14) and even ozone-depleting
species including CClF$_3$ (CFC-13) can be formed, underscoring the
complexity and environmental significance of these degradation
mechanisms.\cite{orr2025atmospheric} Despite these experimental
and modeling advances, the detailed unimolecular decay pathways that
connect energized fluorinated CIs to these stable end products remain
incompletely understood, particularly for reactants R$_1$R$_2$COO with
high initial energy content.\\

\noindent
Computational studies of CI decomposition are challenging for several
reasons. First, the PESs are high-dimensional,
involving multiple shallow minima and low-lying transition states
(TSs).\cite{MM.h2coo:2025,MM.h2coo:2026} Secondly, the reactions often occur under
highly energized conditions, whereby excess internal energy can lead
to non-statistical behavior and pathway branching that is not captured
by minimum-energy reaction paths.\cite{carpenter2005nonstatistical}
Characterizing non-statistical pathways is particularly important
because many coarse-grained simulations and atmospheric chemistry
models rely on RRKM or RRKM-like assumptions when estimating
unimolecular rate constants and product branching ratios. A central
assumption underlying the validity of such approaches is rapid
intramolecular vibrational relaxation (IVR), or more generally
efficient redistribution of internal energy among the available
vibrational modes prior to reaction. For pathways involving multiple
intermediates and transition states with appreciably different
energetics, explicit dynamics simulations are needed to determine
whether the resulting dynamics are statistical or whether direct,
mode-specific, or dynamically biased pathways contribute
significantly. While \textit{ab initio} electronic structure
calculations can characterize stationary points and intrinsic reaction
coordinates, they are generally too expensive to enable long-time,
statistically meaningful molecular dynamics simulations at high levels
of theory.\\

\noindent
Machine-learning-based potential energy surfaces (ML-PESs) emerged as a
promising alternative to statistical unimolecular calculations and
{\it ab initio} MD simulations to address these
challenges.\cite{behler2014representing,fedik2022extending,MM.nnpes:2023}
By learning the underlying PES from electronic structure data, such
models can deliver near \textit{ab initio} accuracy corresponding to
the level of theory of the reference data but at a fraction of their
computational cost. This allows extensive reactive dynamics
simulations to be carried out. Given the importance of the
F-containing Criegee intermediate CF$_3$CHOO and the incomplete
characterization of its decomposition dynamics, the unimolecular
reaction dynamics is investigated. The results provide the population of
fragmentation channels and fundamental mechanistic insight into
generating CHF$_3$ and related products and illustrate the broader
utility of machine-learning-based PESs for studying complex
atmospheric reaction dynamics.\\

\section{Results}

\subsection{Validation of the PES}
It is good practice to train several independent NNs for the same overall data set. In the present work, four such models were trained using PhysNet. Statistical benchmarks for all trained models are  summarized in Table \ref{sitab:evaluation}. All four models achieve
low mean absolute and root mean square errors for both energies and
forces, with $R^2 > 0.999$ for the energy predictions. These results
indicate excellent agreement with the MP2 reference data across the
chemically relevant regions of the PES. Minor variations among the
models reflect differences in training initialization and data
partitioning, but all models are sufficiently accurate for reactive MD
simulations.\\

\begin{figure} [H]
    \centering \includegraphics[width=1.0\linewidth]{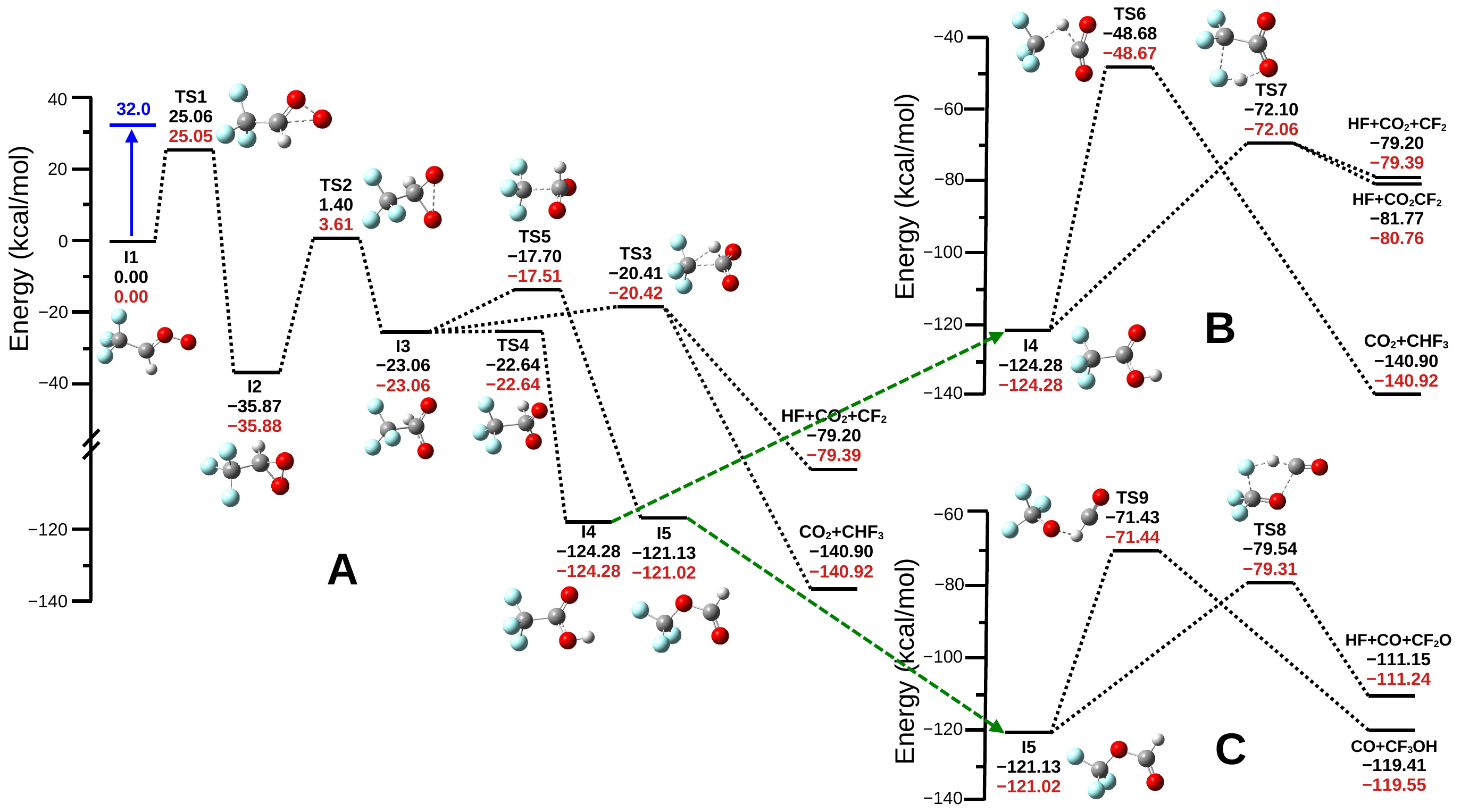}
    \caption{Reaction diagram showing all the reaction pathways
      studied in this work. Panel A: Initial excitation of I1 leading
      to I2 and I3. Panel B: Dynamics following transition from I3 via
      TS4 to I4. Panel C: Dynamics following transition from I3 via
      TS6 to I5. Energies (in kcal/mol) from MP2/aug-cc-pVTZ reference
      calculations (black), and predictions from PhysNet
      (red). Enlarged views of panels A to C with further details are
      shown in Figures \ref{sifig:I1_pes} and \ref{sifig:I3_pes}.}
    \label{fig:reaction_diagram}
\end{figure}

\noindent
The reaction diagrams in Figures \ref{fig:reaction_diagram} and
\ref{sifig:I1_pes} indicate that reactant I1 undergoes a low-barrier
rearrangement ($E^\dagger = 25$ kcal/mol) via TS1 to form I2. The
close agreement between MP2 reference energies and PhysNet predictions
(black vs. red values in Figure \ref{fig:reaction_diagram}) along the
pathways demonstrates that the trained model accurately captures the
overall energetics. Typically, the differences between reference and
the neural network (NN-)trained PES are $\sim 0.1$ kcal/mol for states covering $\sim
150$ kcal/mol, with the exception of TS2. This transition state is
notorious, not only for the present system but also for
H$_2$COO\cite{MM.h2coo:2025} and for other CIs. With the amount of
excess energy available, I2 undergoes O–O bond cleavage through TS2,
leading to the formation of I3. However, I3 is highly unstable and
exists only for $\leq 0.1$ ps before undergoing dissociation to
CO$_2$+CHF$_3$ or HF+CO$_2$+CF$_2$, or further rearrangement to form
I4 (trifluoroacetic acid) or I5 (trifluoroethyl formate).\\

\noindent
Figures \ref{sifig:sampling} and \ref{sifig:corr} report the energy
distribution of the sampling and the correlation between MP2/aVTZ
reference energies and PhysNet predictions for the test set of Model4,
respectively. The near-perfect linear correlation ($R^2 = 1-10^{-4}$)
demonstrates that the model accurately reproduces the global energy
landscape. The narrow distribution of errors (inset of Figure
\ref{sifig:corr}) indicates that large deviations are rare and that
the model remains stable across a wide range of configurations,
including high-energy regions relevant to reactive dynamics.\\

\begin{figure} [h!]
    \centering \includegraphics[width=0.8\linewidth]{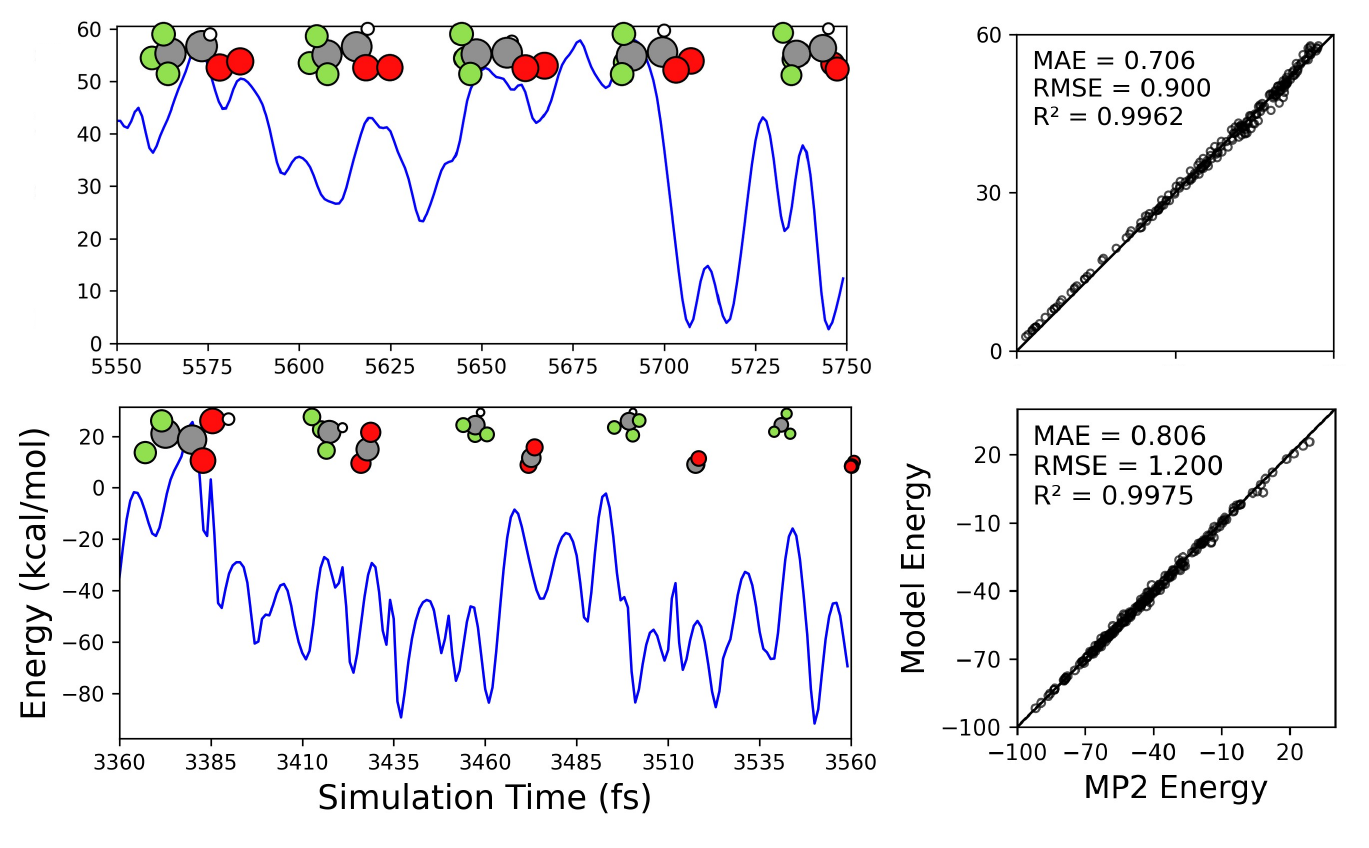}
    \caption{Upper panel shows a representative trajectory for the
      first step in the unimolecular reaction of I1. The reaction
      proceeds through TS1 to form intermediate I2, and subsequently
      through TS2 to yield I3. The lower panel shows one representative trajectory
      shows that I4 can decompose into CO$_2$ + CHF$_3$ via TS6.}
    \label{fig:I1_I2_traj_I4_decom}
\end{figure}

\noindent
In addition to the statistical evaluation of the trained models,
Model4 used in the reactive MD simulations {\it vide infra} was
further evaluated on individual trajectories, see Figure
\ref{fig:I1_I2_traj_I4_decom}. The model reproduces the MP2 energies
with high accuracy. For the upper trajectory, the correlation yields
${\rm MAE}(E) = 0.706$, ${\rm RMSE}(E) = 0.900$, and $R^2 = 0.996$,
whereas for the lower trajectory the corresponding values are 0.806,
1.200, and 0.998, respectively. The upper and lower panels show the
total energy (blue traces) with selected structures for the
I1$\rightarrow$TS1$\rightarrow$I2$\rightarrow$TS2$\rightarrow$I3, and
I4$\rightarrow$TS6$\rightarrow$CO$_2$ + CHF$_3$ transitions,
respectively. In the right hand panels, the correlation between the
reference calculations and the trained ML-PES is reported. The smooth
evolution of the potential energy and the close agreement between MP2
and model predictions along the trajectory confirm that the PES
provides a consistent description of the reaction coordinate. It
should be noted that such evaluations provide a rigorous validation of
the ML-PES on structural ensembles which were not part of the data
sets. The distributions of internal variables sampled from the MD
simulations show a large degree of overlap with those of the training
set, see Figure \ref{sifig:coor}.\\

\subsection{Reactive Dynamics and Fragmentation Channels}
Next, the reaction dynamics of the energized reactant I1 governed by
the PES is characterized. The initial energy distribution of the MD
ensemble is shown in Figure \ref{sifig:initial}. Sampling from an
equilibrium distribution at 300 K ensures that the initial structures
are representative of the vibrational ground state region of I1. The
internal energy content of 32.0 kcal/mol assigned to the reactant is
meaningful (see Methods) and allows to overcome TS1. Such a
preparation mimics the energy content of I1 after ozonolysis of the
initial reactant, i.e. formation of CF$_3$CHOO from ozonolysis of
(1E)-1,3,3,3-tetrafluoropropene (HFO-1234ze(E),
CFH=CH-CF$_3$).\cite{mcgillen:2023} The dependence of the ensuing
dynamics on the initial internal energy content is further probed
below.\\

\noindent
Among the 10000 trajectories, $29.6 \pm 0.6$\% (2957) crossed TS1, of
which $0.6 \pm 0.1$\% (62) became trapped in I2 on the time scale of
the simulations (1 ns), whereas the remaining trajectories ($29.0 \pm
0.6$\%) proceeded through TS2 after 1 ns, and none of them remained in
I3. Once formed, I3 is extremely short-lived (less than 0.1 ps),
consistent with its shallow minimum on the PES, shown in Figure
\ref{sifig:I3_pes}. The MD simulations reveal that, rather than
following a single intrinsic reaction coordinate, I3 connects to
multiple competing decay channels. With the excess energy available
after TS3, formation of CO$_2$ + CHF$_3$ is not guaranteed, contrary
to the prediction from the {\it ab initio} IRC calculation. In
particular, the fate of the hydrogen atom is strongly influenced by
the amount and the internal degree(s) of freedom the excess energy is
contained in, leading either to recombination with carbon to form
CHF$_3$ or abstraction by fluorine to yield HF.\\

\noindent
{\it Dynamics following I3:} Out of the 29.0\% of trajectories that
reach I3, the population is predominantly distributed between the two
pathways leading over TS3 and TS4: 56.1\% trajectories proceed through
TS3 and 38.5\% through TS4, and only 5.4\% continue across TS5. Among
the trajectories passing through TS3, the majority 89.4 \% (1259 out
of 1408) led to CO$_2$ + CHF$_3$, with the remainder forming HF +
CO$_2$ + CF$_2$. Figure \ref{sifig:I2_traj_1} illustrates these two
pathways by showing two representative trajectories.\\

\noindent
In addition to direct dissociation, I3 can rearrange to form
intermediates I4 or I5, see Figure \ref{sifig:I2_traj_2}. These
pathways involve significant structural reorganization and therefore
redistribute the excess energy among different vibrational modes. The
formation of I4 and I5 opens additional reaction channels that
contribute to the overall product diversity observed in the MD
simulations, see Figure \ref{fig:reaction_diagram}B/C, which are
considered next. \\

\begin{figure} [H]
    \centering \includegraphics[width=1.0\linewidth]{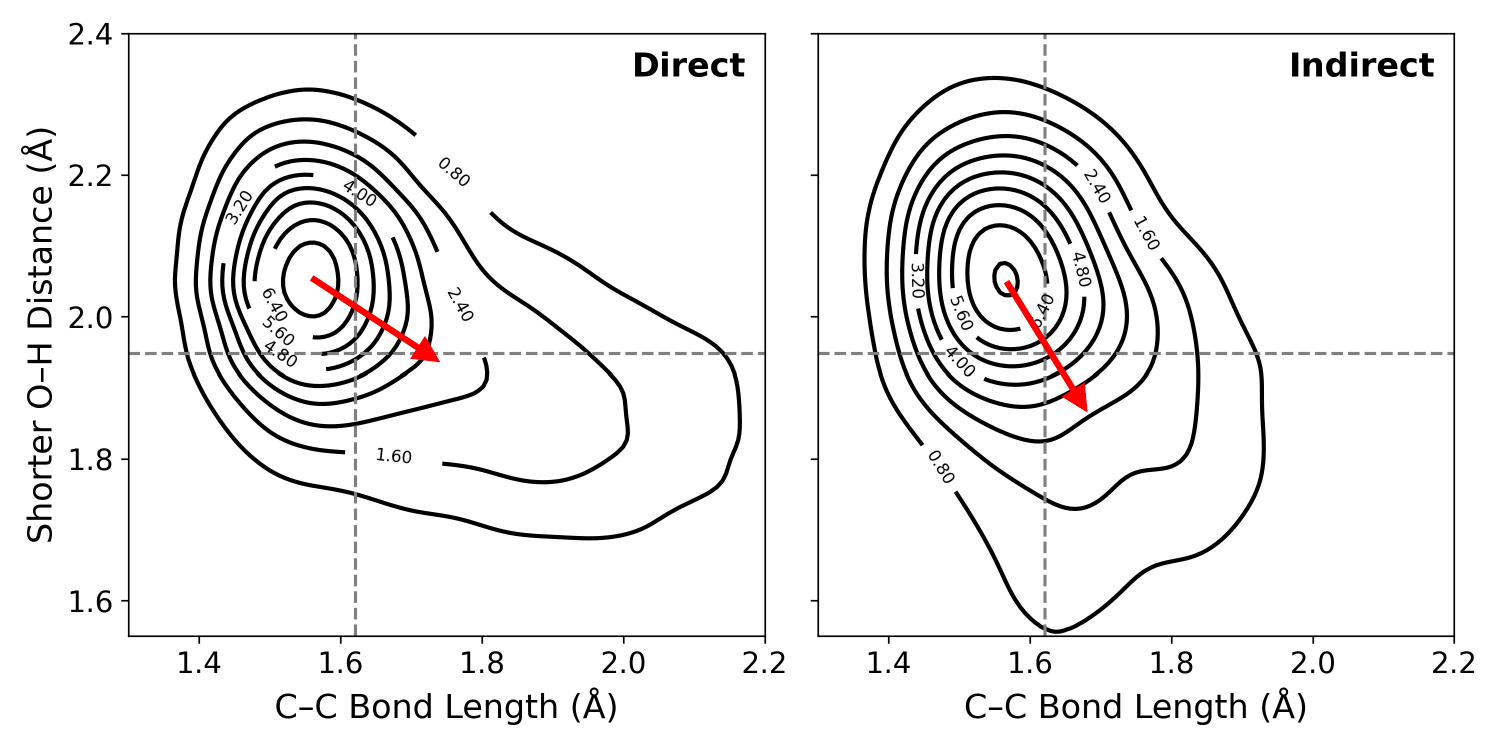}
    \caption{Normalized
      probability distributions $P(r_{\rm CC},r_{\rm OH})$ while
      sampling I3 well for each of the direct (left) and indirect
      (right) channels. For the indirect channel, the C--C stretch (with some admixture of the O--H separation) is the main progression coordinate, whereas the indirect pathway proceeds along a combination of the O--H and C--C bonds. The numbers in the contour lines indicate the
      normalized intensity of geometries. The dashed lines represent
      the equilibrium structure of I3. The red arrows illustrate the
      direction of structural evolution along the corresponding channels.}
    \label{fig:I3_coor_2D}
\end{figure}

\noindent
Figure \ref{fig:I3_coor_2D} shows the normalized probability
distributions $P(r_{\rm CC},r_{\rm OH})$ of the geometries sampled in
the I3 well for the direct and indirect channels. In both cases, the
highest probability density is concentrated close to, but shifted
from, the equilibrium geometry of I3, indicated by the dashed
lines. Relative to TS4, the sampled geometries exhibit elongated C–C
bonds and shortened O–H bonds, consistent with the structural
characteristics of TS3.\\

\noindent
{\it Dynamics through intermediate I4:} Intermediate I4 is connected
to two low-energy exit channels, as shown in Figure
\ref{fig:reaction_diagram}B. Notably, TS7 is readily accessible with
the available excess energy, while the dissociation energy of CC bond
to form HF + CO$_2$ + CF$_2$ is only 1 to 2 kcal/mol. Consequently,
trajectories reaching I4 may either decompose through TS6 to form the
thermodynamically most stable products CO$_2$ + CHF$_3$, or proceed
through TS7, where they either a) proceed through breaking the CC bond
(to form HF + CO$_2$ + CF$_2$), or b) persist as HF + CO$_2$CF$_2$ for
the remainder of the dynamics on the 1 ns time scale. This contrasts
with the single product pathway predicted by IRC calculations and
highlights the importance of explicit dynamical simulations for
determining realistic product state distributions.\\

\noindent
For the trajectories reaching I4, the majority 84.7\% (819 of 967)
passed through TS7. The remaining 15.3\% (148 trajectories) proceeded
through TS6 to yield CO$_2$ + CHF$_3$. Of the trajectories passing
through TS7, 96.2\% (788 out of 819) formed HF + CO$_2$ + CF$_2$,
whereas 3.8\% (31 out of 819) remained as HF + CO$_2$CF$_2$. Figure
\ref{sifig:I4_traj} illustrates the two pathways after TS7: The
CO$_2$CF$_2$ fragment may subsequently dissociate into CO$_2$ + CF$_2$
if sufficient excess energy remains in the relevant degree of freedom
(C--C stretch). Such sequential fragmentation highlights the
importance of treating dissociation dynamics explicitly rather than
assuming immediate formation of the most stable products. The bottom
panel of Figure \ref{fig:I1_I2_traj_I4_decom} shows a representative
trajectory in which I4 decomposes via TS6 to yield CO$_2$ + CHF$_3$.\\

\noindent
{\it Dynamics through I5:} For I5, the PES reveals two accessible
pathways, see Figure \ref{fig:reaction_diagram}. Among the 36
trajectories that reach I5, half proceed via TS8 to form HF + CO +
CF$_2$O, whereas the other half follows TS9, leading to CO +
CF$_3$OH. Figure \ref{sifig:I5_traj} depicts two representative
decomposition pathways of I5.\\

\noindent
{\it Summary:} Table \ref{sitab:yield} summarizes all products that
were identified for the unimolecular decay of internally excited
CF$_3$CHOO from 10000 trajectories. In addition to the dominant
product CO$_2$ + CHF$_3$ (14\%), significant yields of HF-containing
products are also observed. The presence of residual intermediates
such as I2 reflects incomplete reaction on the 1 ns time scale and
highlights the competition between different decay channels under
highly excited conditions. In addition, minor radical-forming channels
included COOH + CF$_3$, CF$_2$ + FCOOH, OH + CO + CF$_3$, H + CO$_2$ +
CF$_3$ and OH + CF$_3$CO. These species are not shown in the reaction
diagram (Figure \ref{fig:reaction_diagram}), and are highlighted in
blue in Table \ref{sitab:yield}. Approximately 70\% of the
trajectories remain in I1 because intramolecular vibrational
redistribution is efficient and crossing TS1 for such trajectories
occurs on time scales longer than 1 ns.\\

\begin{figure} [H]
    \centering \includegraphics[width=1.0\linewidth]{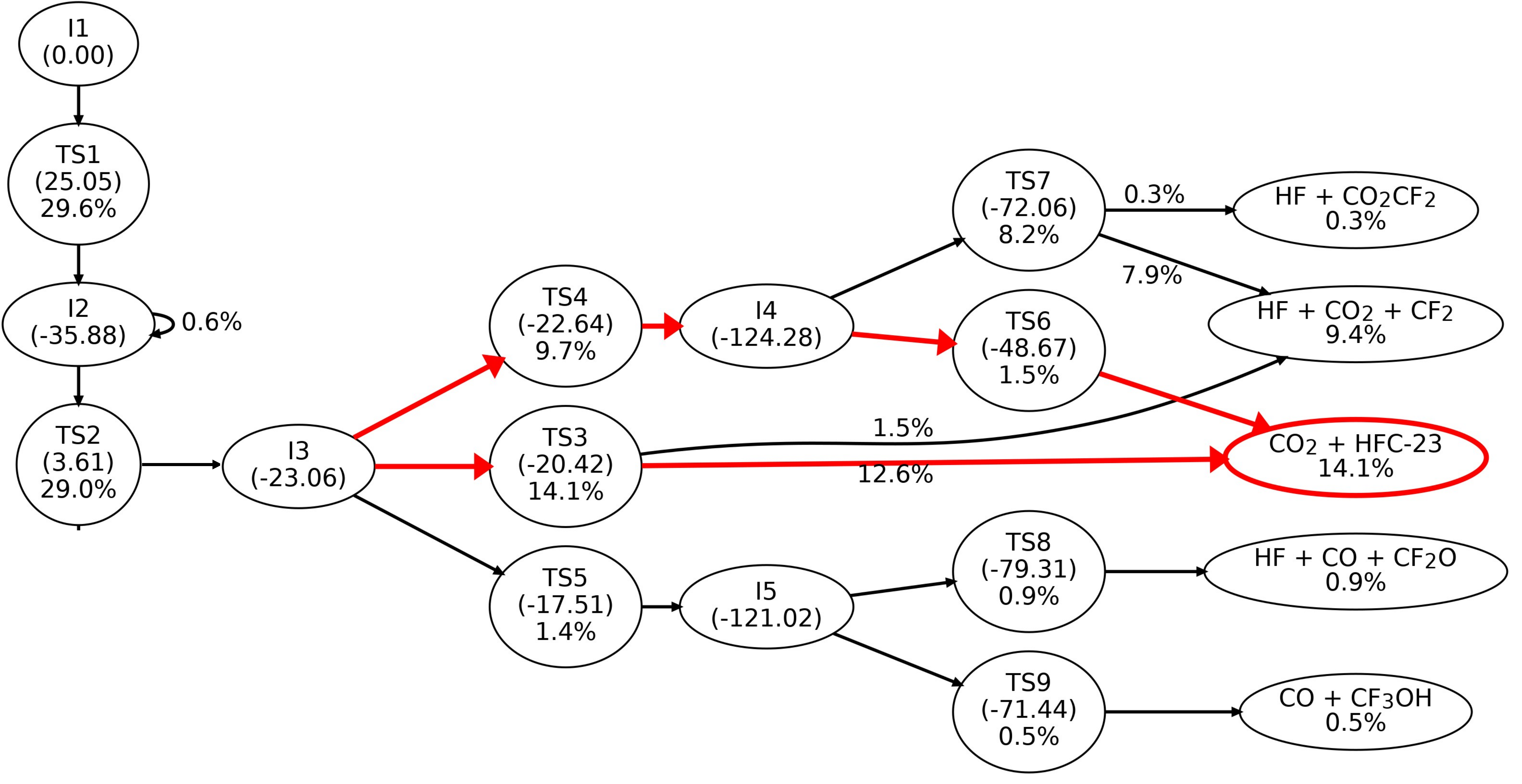}
    \caption{Tree diagram of reaction pathways showing the percentage
      of trajectories (relative to 10000 total trajectories) passing
      through each TS, intermediate, and product channel. Energies are
      given in parentheses (kcal/mol). Percentages quantify the flux
      along individual pathways and provide a compact overview of
      branching dynamics and product formation probabilities.}
    \label{fig:reaction_tree}
\end{figure}

\noindent
A graphical overview is provided in Figure \ref{fig:reaction_tree},
which summarizes the reaction pathways together with the flux between
the species. Trajectories evolve through a series of intermediates
(I2–I5) and across TSs (TS1–TS9), branching into multiple competing
channels and ultimately forming a broad range of product states. The
distribution of trajectories across these pathways highlights the
relative importance of each reaction channel. For the time evolution
of the cumulative populations of the intermediates I4 and I5, as well
as the major products CO$_2$+CHF$_3$ and HF+CO$_2$+CF$_2$, see Figure
\ref{sifig:population_vs_tau}. Note that the percentages of pathways leaving I3 do not sum to the 29\% of trajectories arriving at I3 because only the major channels are shown here. Minor radical-forming channels, including COOH + CF$_3$, CF$_2$ + FCOOH, OH + CO + CF$_3$, H + CO$_2$ + CF$_3$, and OH + CF$_3$CO, are omitted for clarity (see Table \ref{sitab:yield}). \\

\subsection{Analysis of the Decomposition Pathways}
First, the decomposition pathways are individually analyzed. To put
the internal energy distribution $E_{\rm int}$ in each of the reactant
species (I2 to I5) into perspective with the transition state energies
separating them from the respective products, Figure
\ref{sifig:energy} shows the corresponding probability distributions,
$P(E_{\rm int})$. Comparison of the energy distributions with the TS
energies shows that all exit channels are energetically accessible
from their respective intermediates. However, energetic accessibility
alone does not determine the branching ratios. In particular, for I3,
although TS4 is the lowest-energy pathway, more trajectories proceed
through TS3, indicating that the product branching is governed not
only by the total excess energy but also by how that energy is
distributed among the internal degrees of freedom, and how easily and
on what time scale energy flows into the degree(s) of freedom that
lead to the respective FS. This energy flow is largely controlled by
the intramolecular couplings.\\

\begin{figure} [H]
    \centering \includegraphics[width=0.8\linewidth]{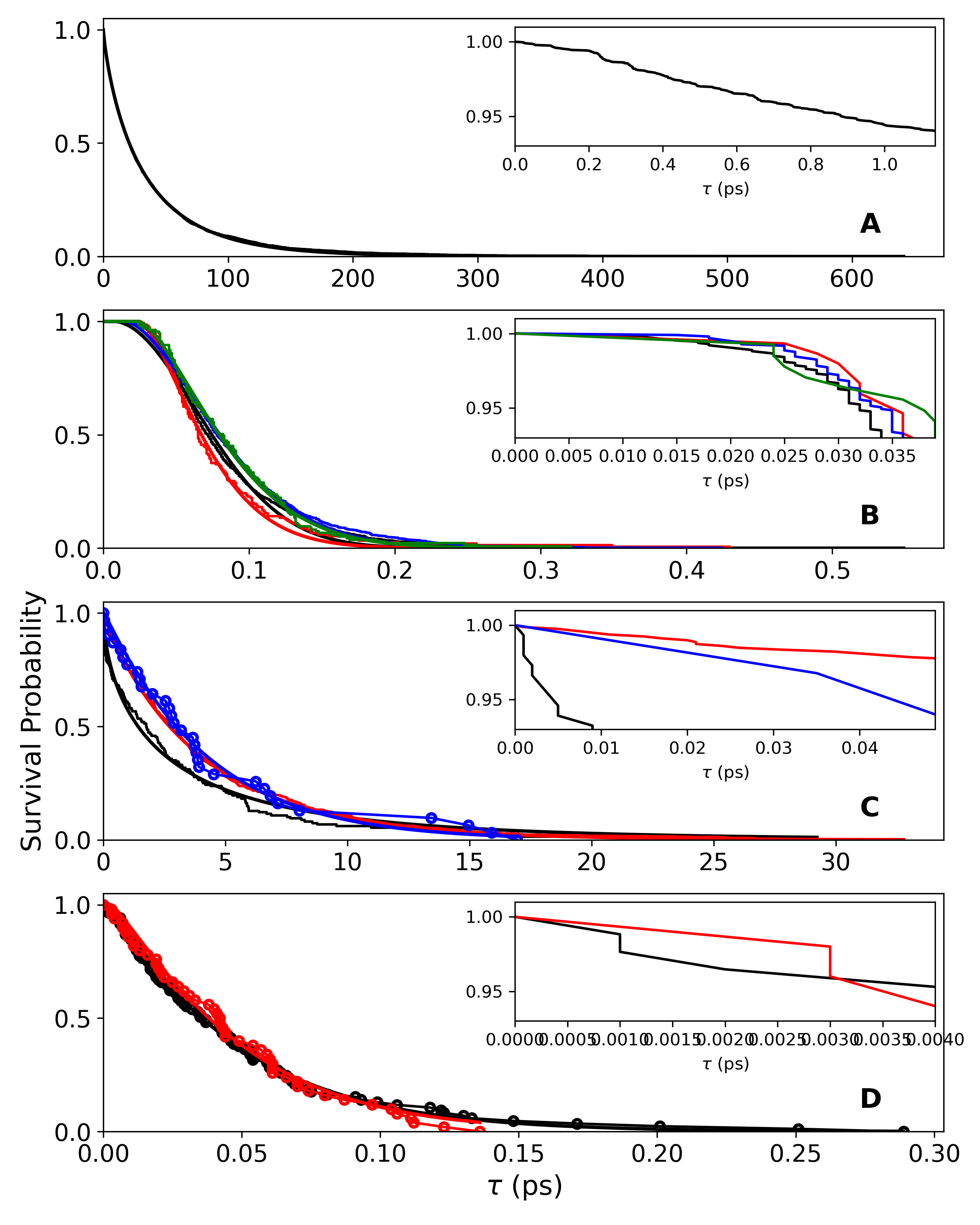}
    \caption{The survival fractions $N(\tau)/N(0)$ for intermediates
      I2 (Panel A), I3 (Panel B), I4 (Panel C), and I5 (Panel D)
      corresponding to the reaction channels as shown in Figure
      \ref{fig:reaction_diagram} to form the indicated
      products. Different colors in each panel represent distinct
      product channels for each intermediate I2 to I5. Panel A:
      I2$\rightarrow$I3 (black); Panel B:
      I3$\rightarrow$[CO$_2$+CHF$_3$ (black), HF+CO$_2$+CF$_2$ (red),
        I4 (blue), I5 (green)]; Panel C:
      I4$\rightarrow$[CO$_2$+CHF$_3$ (black), HF+CO$_2$+CF$_2$ (red),
        HF+CO$_2$CF$_2$ (blue)]; Panel D:
      I5$\rightarrow$[HF+CO+CF$_2$O (black), CO+CF$_3$OH (red)]. The
      step-like curves represent the raw survival statistics obtained
      from 10000 MD trajectories, while the smooth curves correspond
      to fits using a stretched-exponential function, excluding the
      plateau region: $\exp\left[-(\tau / \tau_{\rm
          r})^{\beta}\right]$. The corresponding fitting parameters
      are summarized in Table \ref{tab:fitting}. Note that $x-$axis
      ranges differ for each panel.}
    \label{fig:survival}
\end{figure}

\noindent
Next, the survival fractions $N(\tau)/N(0)$ for each reactant state
with respect to the corresponding final state are reported, see Figure
\ref{fig:survival}. The Kohlrausch (or stretched-exponential) function
$\exp\left[-(\tau / \tau_{\rm r})^{\beta}\right]$ with relaxation time
$\tau_{\rm r}$ and exponent $\beta$ was originally introduced to
describe relaxation processes that deviate from simple exponential
behavior and cannot be represented in a meaningful fashion by a finite
sum of exponential decays over extended time
scales.\cite{kohlrausch:1854,watts:1970} Such behavior is generally
associated with a continuous distribution of relaxation times or
distributed energy barriers, for which the stretched exponential
provides a compact phenomenological
description.\cite{austin:1974,austin:1975,richert:2002} While it also
offers a convenient alternative to multi-exponential fitting, its
greater value lies in the physical interpretation of the exponent,
$\beta$, as an indicator of dynamical heterogeneity.\\

\begin{table}[H]
\centering
\begin{tabular}{llrccc}
\hline
Initial & Final & $N$ & $\tau_0$ (ps) & $\tau_{\rm r}$ (ps) & $\beta$ \\
\hline
I2 & I3 & 2895 & 0.023 & $32.56 \pm 0.79$ & $0.82 \pm 0.01$ \\
\hline
I3 & CO$_2$+CHF$_3$ & 1259 & 0.007 & $0.08 \pm 0.00$ & $1.93 \pm 0.05$ \\
I3 & HF+CO$_2$+CF$_2$ & 149 & 0.025 & $0.05 \pm 0.00$ & $1.48 \pm 0.11$ \\
I3 & I4 & 967 & 0.015 & $0.08 \pm 0.00$ & $1.65 \pm 0.05$ \\
I3 & I5 & 135 & 0.024 & $0.07 \pm 0.00$ & $1.52 \pm 0.11$ \\
\hline
I4 & CO$_2$+CHF$_3$ & 148 & 0.001 & $2.58 \pm 0.34$ & $0.62 \pm 0.07$ \\
I4 & HF+CO$_2$+CF$_2$ & 788 & 0.001 & $4.02 \pm 0.18$ & $0.93 \pm 0.03$ \\
I4 & HF+CO$_2$CF$_2$ & 31 & 0.035 & $4.21 \pm 0.83$ & $1.03 \pm 0.25$ \\
\hline
I5 & HF+CO+CF$_2$O & 85 & 0.001 & $0.05 \pm 0.01$ & $1.07 \pm 0.12$ \\
I5 & CO+CF$_3$OH & 50 & 0.003 & $0.05 \pm 0.01$ & $1.15 \pm 0.19$ \\
\hline
\end{tabular}
\caption{Stretched-exponential fit parameters of survival
  probabilities for different species in each reaction channel. $N$ is
  the number of raw data and $\tau_0$ is the plateau time. Values of
  $\beta \approx 1$ indicate near single-exponential decay, consistent
  with statistical unimolecular kinetics in the RRKM limit. For $\beta
  > 1$, the dynamics exhibit compressed-exponential behavior,
  suggesting deviations from RRKM-type kinetics. Error bars were
  obtained via bootstrapping. Clear non-RRKM behavior is observed for
  I3, particularly in its decomposition into CO$_2$ + CHF$_3$ and HF +
  CO$_2$ + CF$_2$ (Panel A). In contrast, I4 and I5 show near-RRKM
  behavior (Panels B and C). One special case is I4 decomposing into
  CO$_2$ + CHF$_3$, where an exceptionally rapid decay is observed.}
\label{tab:fitting}
\end{table}

\noindent
Several raw survival curves, $N(\tau)/N(0)$, exhibit a short-time
plateau prior to the onset of decay, particularly pronounced in Panel
B of Figure \ref{fig:survival}, which can be interpreted as an
apparent initialization period. The intermediates I3 to I5 exhibit
rather different relaxation times $\tau_{\rm r}$, see Table
\ref{tab:fitting}. The corresponding lifetimes of the states leading
to the different products are shown in Figure \ref{sifig:life_kde}.
Largely independent on the product, the relaxation times of the I3 and
I5 states are on the sub-100 fs time scale. All decays are
characterized by $\beta > 1$, hence compressed exponential behaviour
is found for the dynamics leaving these two states. On the other hand,
for the I4 state the $\tau_{\rm r}$ are on the ps-time scale and
differ by almost a factor of two depending on the target
channel. Formation of CO$_2$+CHF$_3$ is associated with $\tau_{\rm r}
\sim 2.5$ ps whereas access to the HF-forming channels is
characterized by $\tau_r = 4.0$ and $\tau_r = 4.5$ ps,
respectively. The Kohlrausch exponent $\beta \sim 1$ or smaller,
i.e. the dynamics is stretched exponential. The lifetimes in each of
the intermediates do not correlate with the barriers separating them
from their corresponding products. One example is I5 for which $\tau_r
= 0.05$ ps but the barriers separating the two products differ by 8
kcal/mol: the pathway across TS8 features a barrier height of 42
kcal/mol, whereas that across TS9 is 50 kcal/mol.\\

\noindent
These MD simulations reveal that the energy profile (Figure
\ref{fig:reaction_diagram}) is not a useful proxy to infer reaction
kinetics. For instance, entropic contributions, dynamical bottlenecks,
or restricted transition pathways could effectively trap the system in
the I4 state, prolonging its lifetime. It is also noted that the
lifetimes do not exhibit any clear correlation with the corresponding
product yields of the individual exit pathways. This suggests that the
decay of a given intermediate is not strongly biased toward a
particular channel. In other words, once the system reaches an
intermediate, the probability for exiting through a specific pathway
is not governed by how long the system resides in that state. Instead,
the branching into different products is likely controlled by subtle
dynamical factors at or beyond the transition regions, such as the
topology of the dividing surfaces or momentum distributions.\\

\section{Discussion and Conclusion}
This study characterized, in a quantitative fashion, the possible
reaction products of the energized CF$_3$CHOO Criegee intermediate
following unimolecular reaction dynamics. A particular focus was on
generating the HFC-23 (CHF$_3$), which has a very high global warming
potential.\\

\noindent
Key to the present work was a highly accurate representation of the
MP2-reference data through a NN-PES. The PhysNet architecture provides
a global, reactive PES suitable for dynamical studies. T1 diagnostics
indicate, that for most intermediates and transition states a
single-reference treatment is adequate (Table \ref{sitab:T1}). One of
the exceptions is TS1, for which the T1-diagnostic points towards
increased multireference character. As the energy of this transition
state is lower by $\sim 2$ kcal/mol at the CASPT2 level (Table
\ref{sitab:caspt2}) compared with the value from MP2 calculations, the
rates obtained from the present work are lower limits. Including
multireference effects in constructing the NN-PES is expected to yield
somewhat faster reaction dynamics and enhanced reactivity. In
addition, the close agreement of vibrational frequencies obtained from
{\it ab initio} calculations and the NN-PES confirms that the PES
accurately captures the local shape of the potential near the saddle
points, shown in Table \ref{sitab:freq}.\\

\noindent
It is also possible to compare the present results with experimental
measurements of CHF$_3$ (HFC-23) formation following ozonolysis of
HFO-1234ze(E).\cite{orr2025atmospheric} The present simulations
predict a CHF$_3$ formation probability of $\sim 14$ \% from the
decomposition of the CF$_3$CHOO Criegee intermediate, compared with an
experimental yield of $(7.9^{+0.4}_{-0.2})$ \% from the overall
ozonolysis reaction. It is useful to briefly juxtapose the two
approaches. Experiments were carried out with a collisional
environment present (N$_2$, O$_2$, other species) that can take up
internal energy from the internally excited parent compound CF$_3$CHOO
which tends to lower the rate. Furthermore, in the experiments the
branching from the primary ozonide into the two competing Criegee
intermediate channels is unknown: ozonolysis of 3,3,3-trifluoropropene
can either form 2,2,2-trifluoroacetaldehyde oxide + formaldehyde or
2,2,2-trifluoroacetaldehyde + formaldehyde oxide. The branching ratio
for the two pathways can not be measured directly and depending on the
assumptions made, the amount of HFC-23 differs. On the other hand, the
present simulations are carried out in gas phase and the exact amount
of internal energy after ozonolysis is unknown. Overall, it is found
that the present simulations and recent measurements agree
qualitatively and find a minor channel ($\sim 10$ \%) to form HFC-23
which is a species with important consequences for the chemical
development of Earth's atmosphere. The time scale of the ML-MD
simulations (1 ns) is meaningful, because it corresponds to typical
collision frequencies of $10^9/{\rm s}$ in the
troposphere\cite{techreport:1976} where CIs are most prevalent.\\

\noindent
Reactive MD simulations on this surface reveal a rich reaction network
involving multiple intermediates and competing decay pathways, see
Figure \ref{fig:reaction_tree}.  While minimum-energy reaction paths
suggest a dominant route toward HF + CO$_2$CF$_2$, the dynamics paint
a more complex picture. Excess internal energy allows the system to
access alternative channels, including CO$_2$ + CHF$_3$, HF + CO$_2$ +
CF$_2$, HF + CO + CF$_2$O, CO + CF$_3$OH, and HF + CO$_2$CF$_2$,
leading to a diverse set of products. The computed product yields
indicate that CO$_2$ + CHF$_3$ is a major but not exclusive outcome of
CF$_3$CHOO unimolecular decay under the simulated conditions. The
almost equal presence of HF + CO$_2$ + CF$_2$ and other minor channels
underscores the importance of non-statistical effects and incomplete
energy redistribution. These findings have direct implications for
atmospheric modeling, where simplified reaction schemes may
underestimate the diversity of products formed from fluorinated CIs.\\

\noindent
Table \ref{sitab:channels} summarizes all molecular fragmentation
channels, i.e. excluding those that contain single H-, C-, or
F-atoms. It is found that the present simulations are capable of
identifying most low-energy products. The only available channels that
were not accessed in the dynamics simulations are channels 37, 52, and
53. Formation of CHF$_2$ along channels 52 and 53 involves replacement
of one F-atom on the CF$_3$-group by a hydrogen atom originating from
the other carbon center. This process would require a substantial
rearrangement (concerted atom replacement between two neighboring
carbon centers) of the molecular framework, making this pathway
unlikely. On the other hand, among the exothermic HF-forming channels
(33 to 38), channel 37 is least favoured energetically.\\

\noindent
More broadly, this work demonstrates the power of machine-learning
PESs to bridge the gap between high-level electronic structure theory
and realistic gas-phase reaction dynamics. Long-time, and
statistically significant numbers of MD simulations on high-level {\it
  ab initio-}based, ML-represented PESs (here MP2, in other cases
CASPT2 and CCSD(T) levels) provide molecular-level insight and
mechanistic information that is otherwise unattainable. Beyond the
immediate relevance of the results that quantify the amount of HFC-23
generated through unimolecular decomposition of CF$_3$CHOO, the
present work provides a generalizable, bottom-up strategy for
designing ``quantum nanoreactors''.  As was demonstrated, most of the
accessible fragmentation channels have already been characterized from
explicit dynamics simulations in the present work (see Tables
\ref{sitab:yield} and \ref{sitab:channels}). Additional product
channels that may be of relevance or interest can be incorporated in a
seamless fashion. Also, it is possible to transfer learn\cite{MM.tl:2022} parts or the
entirety of the reactive NN-representation to yet higher levels of
theory or rigor, such as CCSD(T) or to include multi-reference effects
wherever needed. With this, chemical reaction and decomposition
networks can be generated and investigated at a sufficiently high
quantum chemical level for a molecular-level understanding of
particular parent species or combinations thereof.\\

\noindent
In conclusion, the PhysNet PES developed here provides a detailed and
dynamically accurate description of the unimolecular decay of
CF$_3$CHOO. The insights gained from this study advance our
understanding of fluorinated CI chemistry and highlight the essential
role of machine-learning methods in modern theoretical reaction
dynamics. Future extensions can incorporate collisional energy
transfer or explicit interactions with atmospheric bath gases to
further refine predictions under tropospheric conditions.\\

\section{Methods}
All minima and TSs involved in the unimolecular decay of CF$_3$CHOO
were initially optimized at the MP2/aVTZ level of
theory.\cite{dunning89BtoNe,helgaker1997basis} Intrinsic reaction
coordinate (IRC) calculations were then conducted to verify that each
TS correctly connects the corresponding minima.\\

\noindent
Normal mode sampling was used to generate geometries around each
minimum and TS.\cite{behler2011atom} For the possible decomposition
products, geometries were generated using the Molecular Orbital
PACkage (MOPAC).\cite{stewart2007optimization} Additionally, the
extended tight-binding (XTB) method was employed to generate
geometries starting from each minimum.\cite{bannwarth2019gfn2} The
dataset also includes geometries from intrinsic reaction coordinate
(IRC) calculations. In total, energies and forces for an initial data
set comprising 16,409 structures were computed at the MP2/aVTZ-level
of theory using the MOLPRO suite of programs.\cite{molpro:2020}
Finally, structures with excessively high energies (100 kcal/mol above
the reactant) were removed, resulting in a first dataset consisting of 15781
structures with corresponding energies and forces.\\

\noindent
First, four independent models were trained using the PhysNet neural network architecture to
construct the machine-learned potential energy surface
(ML-PES) whereby the dataset was split into approximately 90/5/5 training/validation/test structures.\cite{MM.physnet:2019} Using all four
PhysNet models, ML-MD simulations were performed at different
temperatures (800 K to 32000K) starting from each minimum (I1 to I5). Then the
geometries along those MD trajectories were evaluated by the four ML
models. If the energy difference between the four models exceeded 1.0
kcal/mol, the corresponding geometry was saved and quantum chemical
calculations were performed. The structure, energy and forces were
then added to the dataset and the four models were retrained. This
procedure was repeated 20 times until the ML-PESs were stable and
reliable. The final dataset comprised 41252 samples which was used in the final training.\\

\noindent
The energy distribution of the final sampling is presented in Figure
\ref{sifig:sampling}. The broad energy coverage of the final dataset
ensures that the trained PESs accurately describe not only the
low-energy stationary points but also the high-energy regions visited
which are sampled during reactive molecular dynamics. Importantly, the
dataset spans both bound regions and dissociative channels, which is
essential for capturing bond-breaking and bond-forming events in
unimolecular decay processes.\\

\noindent
For the ML-MD simulations, the initial structures were generated from
an equilibrium ensemble of I1 at 300 K. Initial velocities were
assigned according to a Maxwell–Boltzmann distribution at 16000 K,
corresponding to a total internal energy of $\sim 32$ kcal/mol and
each trajectory was run for at most 1 ns or until dcomposition
products were formed. The time scale of 1 ns corresponds to a typical
collision time in the stratosphere which is $10^9$
s$^{-1}$.\cite{techreport:1976} The energy released following the
reaction HFO-1234ze(E) + O$_3$ $\rightarrow$ CF$_3$CHOO + HCHO
reactant is $\sim 70$ kcal/mol.\cite{mcgillen:2023} Hence, assigning
approximately half that value to CF$_3$CHOO is a meaningful, albeit
approximate procedure. To quantify the dependence of this choice,
additional simulations were run with somewhat lower (28.0 kcal/mol)
and higher (35.0 kcal/mol) amounts of internal energy. With 28.0
kcal/mol, fewer than 10 \% of trajectories cross TS1 on a time scale
of 1 ns.\\

\section*{Supporting Information} 
The supporting material includes additional validations of the NN-PES,
Figures and Tables.

\section*{Data Availability} 
The reference data that allow to reproduce the findings of this study
are openly available at \url{https://github.com/MMunibas/cf3choo}. \\

\section*{Acknowledgment}
This work was partially supported by the Swiss National Science
Foundation through grants 200021-188724, the NCCR MUST (to MM), the
University of Basel. The authors thank Prof. A. Orr-Ewing for
scientific exchange, in particular pertaining to the measurements.\\


\clearpage

\renewcommand{\thetable}{S\arabic{table}}
\renewcommand{\thefigure}{S\arabic{figure}}
\renewcommand{\thesection}{S\arabic{section}}
\renewcommand{\d}{\text{d}}
\setcounter{figure}{0}  
\setcounter{section}{0}  
\setcounter{table}{0}

\newpage

\noindent
{\bf SUPPORTING INFORMATION: Reaction Pathway Detection using Machine-Learned Energy
  Potentials - Decomposition of Energized CF$_3$CHOO}

\section{Further Validation of the NN-PES}

\noindent
{\it Harmonic Frequencies:} In addition to energetic accuracy, the
ability of the PES to reproduce curvature along reaction coordinates
is critical for reliable dynamics. Table \ref{sitab:freq} compares the
imaginary frequencies of TSs obtained from MP2 calculations and from
the PhysNet PES. The close agreement confirms that the PES accurately
captures the local shape of the potential near the saddle
points. To the best of the authors' knowledge, no measured spectra for
direct comparison are available from infrared and/or Raman
spectroscopy.\\

\noindent
{\it Multi-reference effects:} To further examine the reliability of
the barrier height associated with TS1, which exhibits the largest T1
diagnostic among all stationary points listed in Table \ref{sitab:T1},
CASPT2 calculations were performed using different active spaces and
basis sets. As shown in Table \ref{sitab:caspt2}, the resulting
barrier heights are consistently about 2 kcal/mol lower than the MP2
values, with only minor dependence on the chosen active space or
basis. This systematic shift suggests that MP2 slightly overestimates
the TS1 barrier but still provides a qualitatively correct description
of the reaction energetics. Consequently, the TS1-related reaction
rates obtained from the MP2-based PES should be regarded as
conservative estimates, since a CASPT2-quality PES would likely
predict somewhat enhanced reactivity and faster reaction
dynamics. Given the small magnitude of the correction, MP2-level data
were nevertheless deemed adequate for training the machine-learning
PES.\\

\section{Figures}

\begin{figure} [H]
    \centering \includegraphics[width=0.8\linewidth]{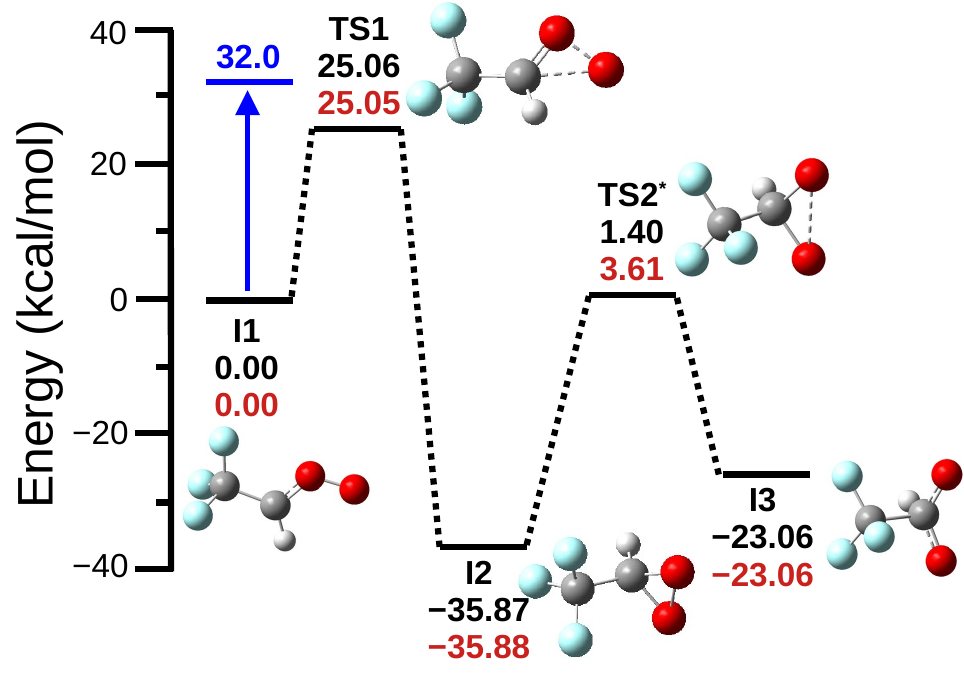}
    \caption{The initial velocities are assigned according to the
      Maxwell–Boltzmann distribution, which represents the statistical
      distribution of atomic speeds in a system at thermal equilibrium
      at 32.0 kcal/mol. Black values correspond to MP2/aug-cc-pVTZ
      reference data, while red values denote predictions from the
      PhysNet model. Note that brute-force {\it ab initio} methods,
      including QST2 or QST3 calculations, were unable to locate TS2,
      whereas using the PhysNet-PES successfully identifies it,
      highlighting one advantage of a NN-based representation. The
      reported MP2 energy is obtained from a single point {\it ab
        initio} calculation performed on the structure optimized by
      PhysNet.}
    \label{sifig:I1_pes}
\end{figure}

\begin{figure} [H]
    \centering \includegraphics[width=0.8\linewidth]{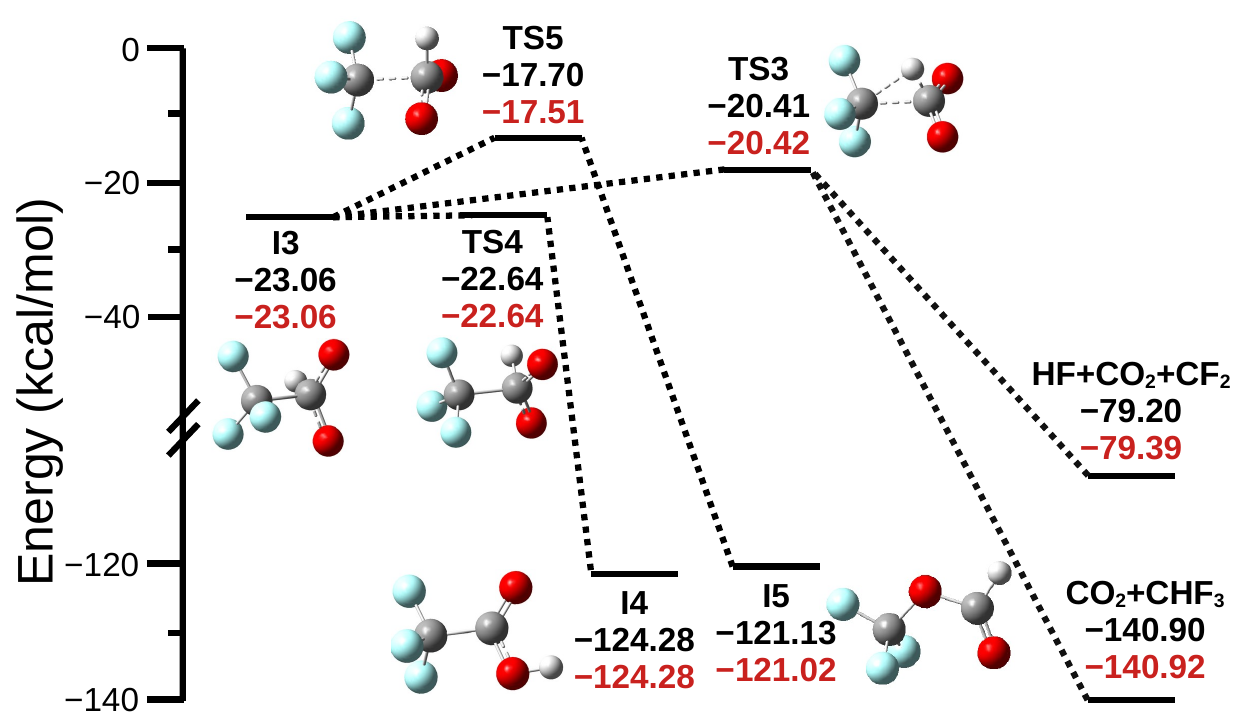}
    \caption{Reaction steps following formation of I3. This
      intermediate either decomposes directly via TS3, or
      alternatively isomerize to I4 (via TS4) or I5 (via TS5). Black
      and red numbers refer to MP2/aug-cc-pVTZ reference data and
      predictions from PhysNet.}
    \label{sifig:I3_pes}
\end{figure}

\begin{figure} [H]
    \centering \includegraphics[width=0.8\linewidth]{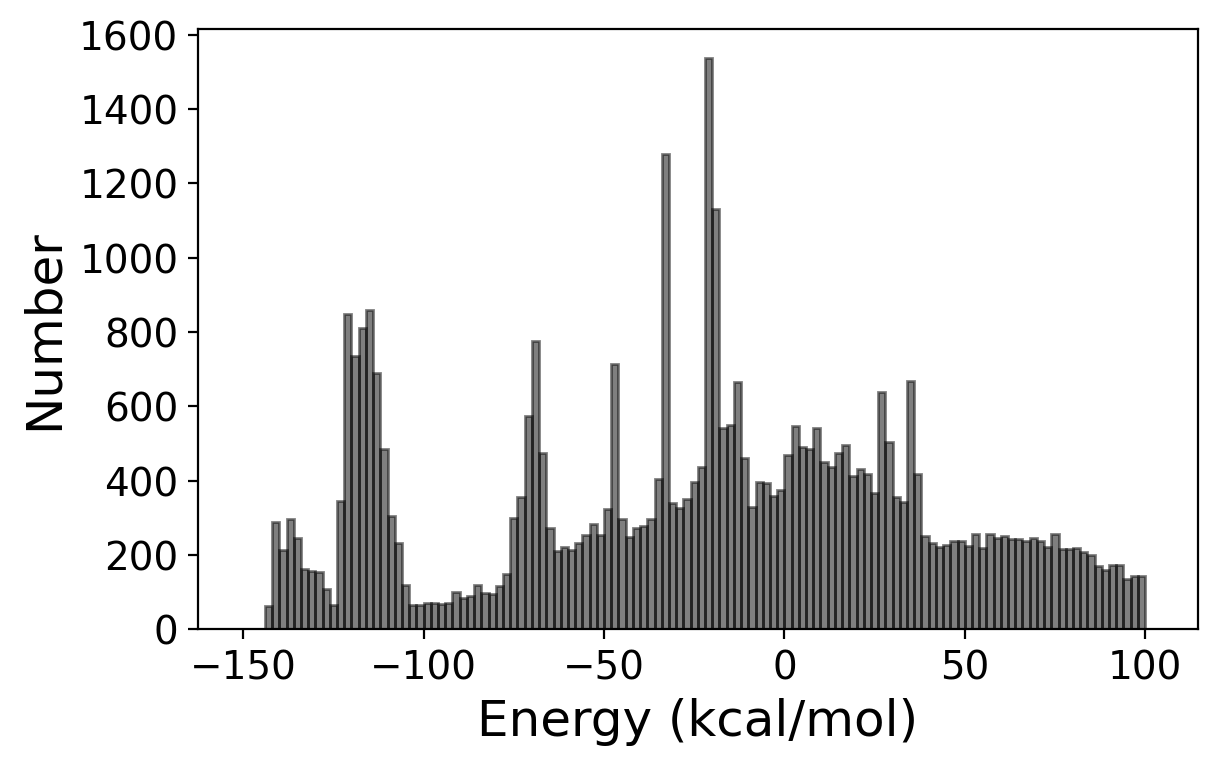}
    \caption{Energy distribution obtained from the final sampling. The
      global minimum, corresponding to the product CO$_2$ + CHF$_3$
      (HFC-23), is located at -140 kcal/mol relative to the
      reactant. Structures with excessively high energies (greater
      than 100 kcal/mol above the reactant) were excluded from the
      dataset.}
    \label{sifig:sampling}
\end{figure}

\begin{figure} [H]
    \centering
    \includegraphics[width=0.8\linewidth]{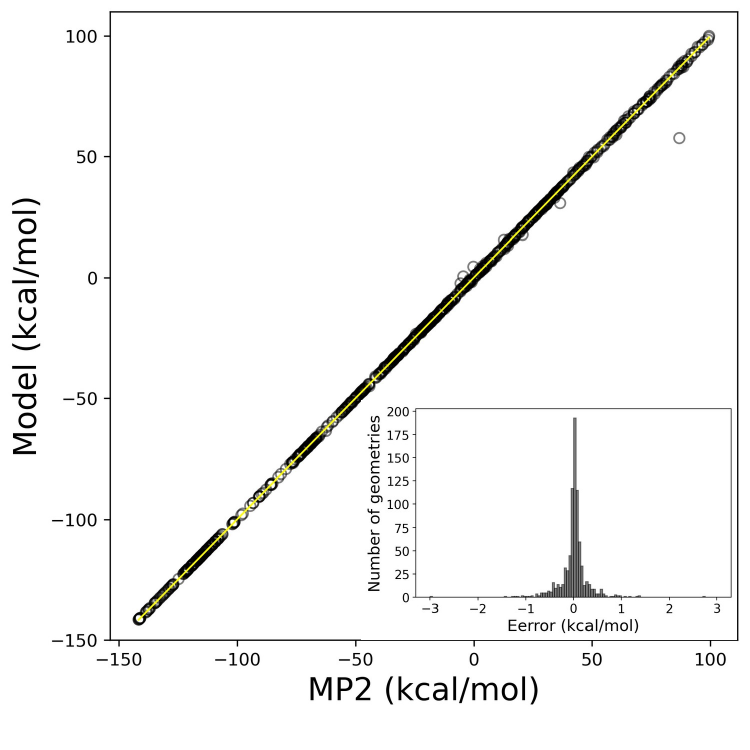}
    \caption{Correlation between MP2/aVTZ reference energies and
      Model4 predictions for the test dataset. The yellow line
      represents perfect 1:1 agreement. The inset displays the error
      distribution. The ten largest outliers correspond to four
      CF$_3$+CHO$_2$ structures, three OH+CF$_3$CO structures, and
      three twisted I2/I3 structures.}
    \label{sifig:corr}
\end{figure}

\begin{figure} [H]
    \centering \includegraphics[width=0.8\linewidth]{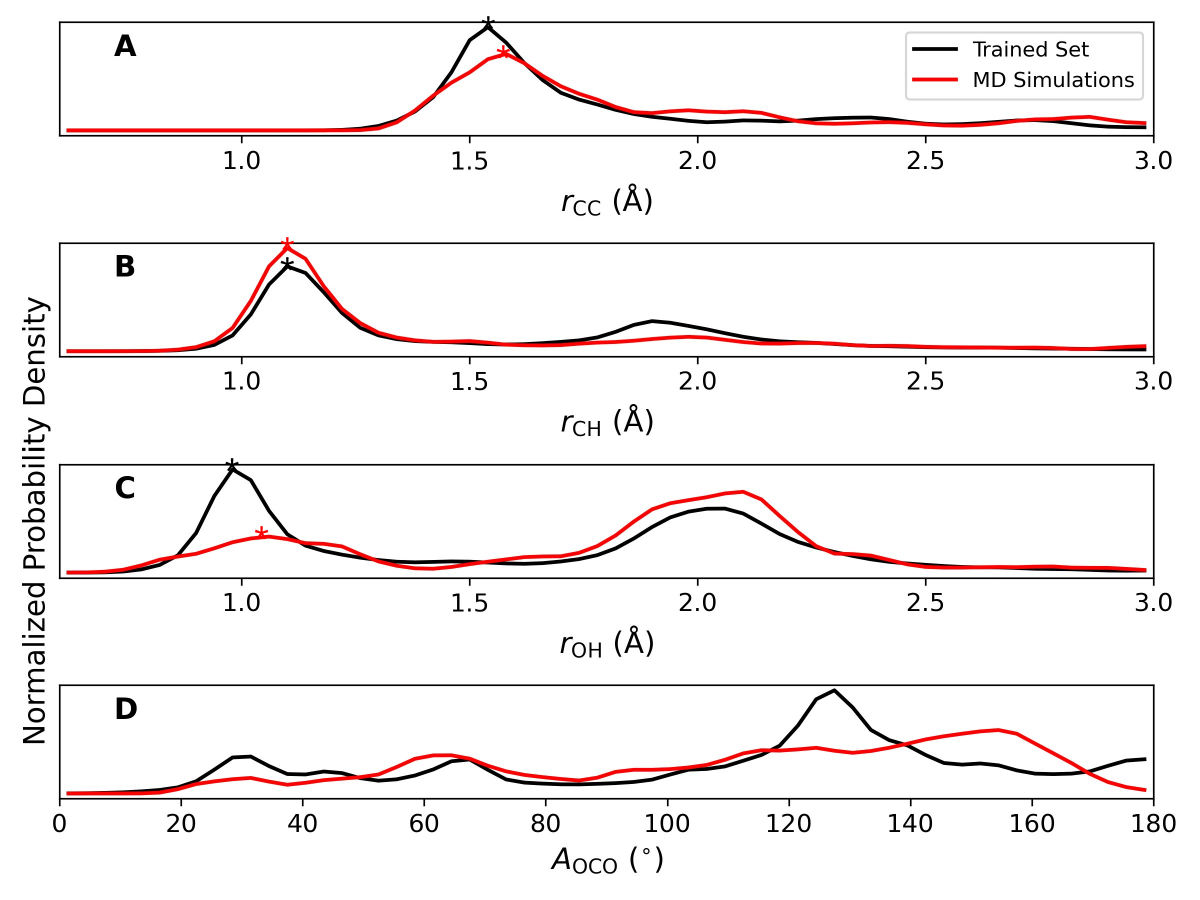}
    \caption{Comparison between the composition of the training set
      and structures sampled during 20 representative MD simulations
      that span the entire region of the PES; all ranges of the
      respective coordinates are represented adequately in the
      training set; i.e. there are no ``uncovered ranges'' or obvious
      "holes". The main peaks (indicated by stars) in Panels A, B, and
      C relate to the equilibrium bond lengths of the respective
      bonds. The secondary peaks in Panels B and C arise from the fact
      that the hydrogen atom can form either a C-H or an O–H bond. For
      example, when an O–H bond is formed, the C–H distance increases,
      giving rise to the second peak in Panel B; conversely, when a
      C–H bond is formed, a corresponding second peak appears in Panel
      C. The O–C–O angle spans a broad range from approximately
      20$^{\circ}$ to 180$^{\circ}$, depending on the molecular
      species.}
    \label{sifig:coor}
\end{figure}

\begin{figure} [H]
    \centering \includegraphics[width=0.8\linewidth]{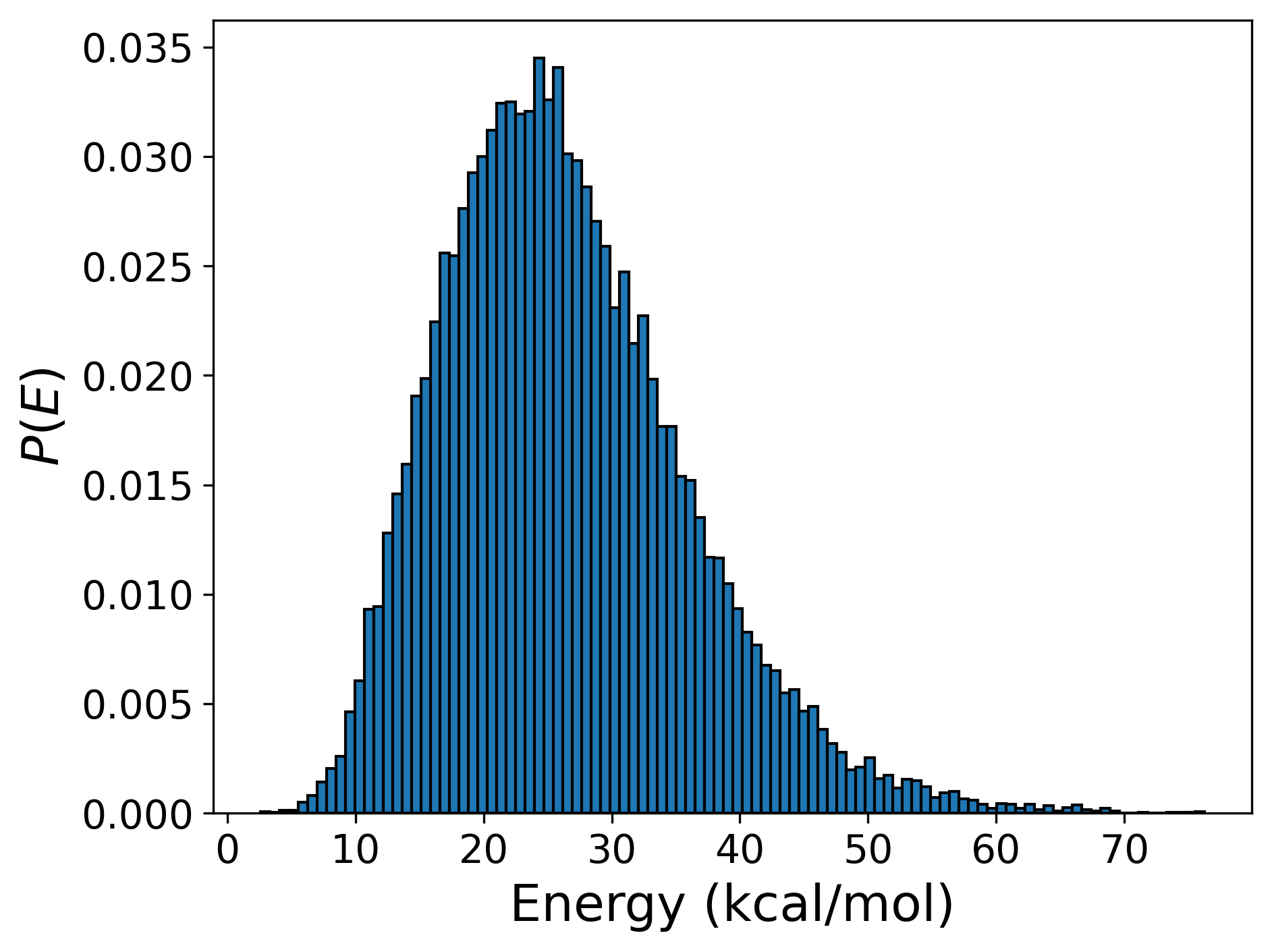}
    \caption{The initial energy distribution of the 10000 independent
      simulations, which reflects the thermal conditions imposed at
      the start of the trajectories. The initial structure is obtained
      from an equilibrium sampling of I1 at 300 K, yielding potential
      energies ranging from 0.07 to 0.72 kcal/mol. The initial
      velocities are assigned according to the Maxwell–Boltzmann
      distribution, which represents the statistical distribution of
      atomic speeds in a system at thermal equilibrium at 32.0 kcal/mol.}
    \label{sifig:initial}
\end{figure}

\begin{figure} [H]
    \centering \includegraphics[width=0.8\linewidth]{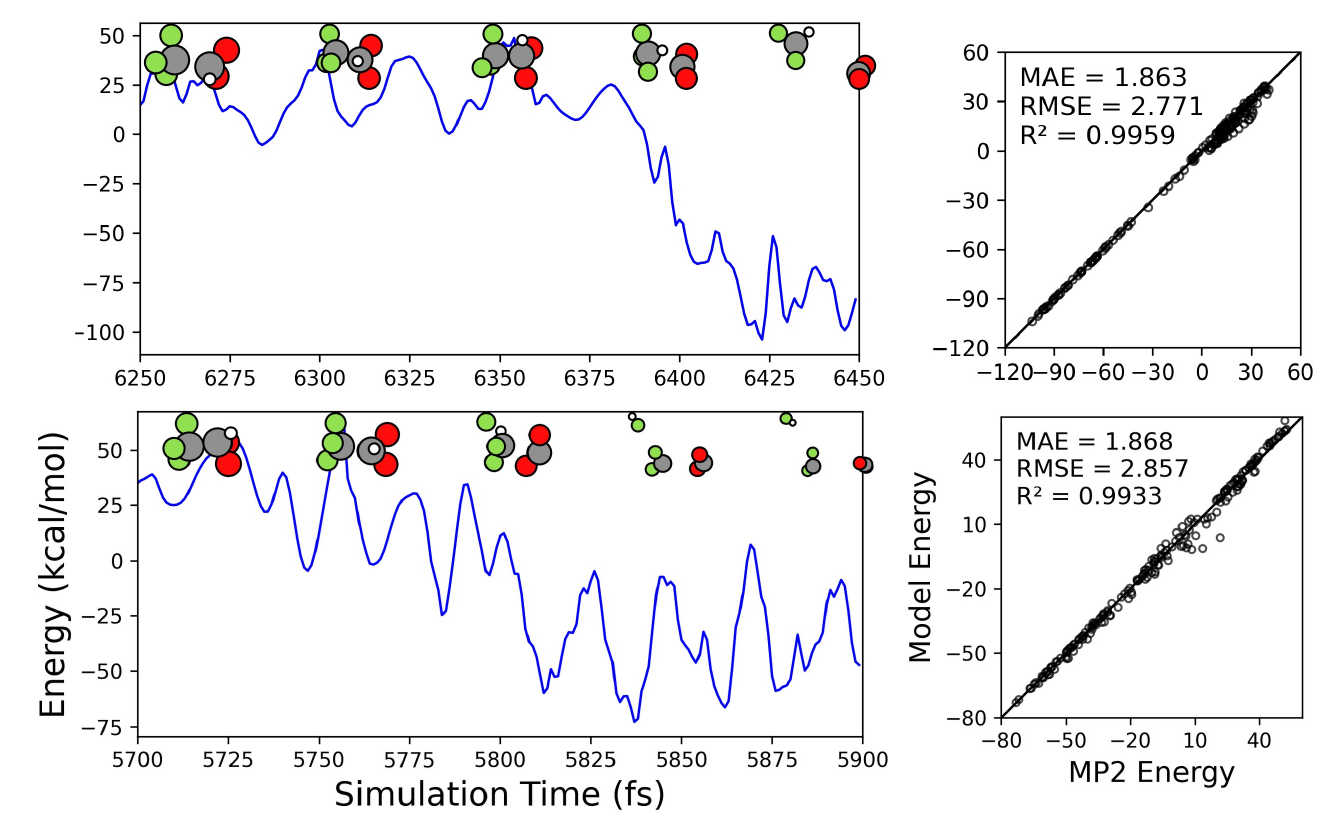}
    \caption{Two representative trajectories illustrating the
      decomposition of I3 are shown. In the first pathway (upper
      panel), the reaction proceeds through TS3 to form CO$_2$ +
      CHF$_3$. In an alternative pathway (lower panel), the H–F bond
      forms instead of the C–H bond, resulting in the formation of HF
      + CO$_2$ + CF$_2$.}
    \label{sifig:I2_traj_1}
\end{figure}

\begin{figure} [H]
    \centering \includegraphics[width=0.8\linewidth]{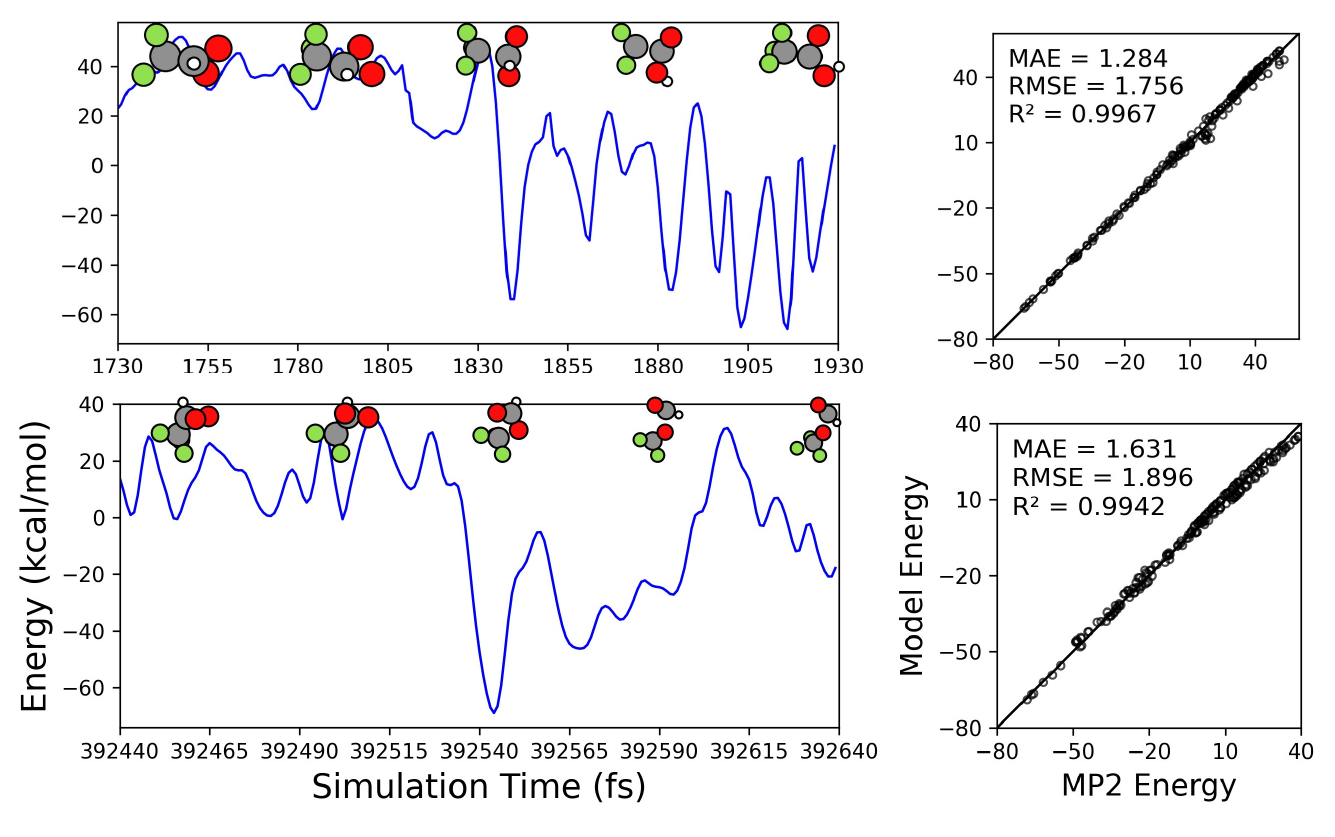}
    \caption{Two representative trajectories illustrate that I3 can
      either convert to I4 via TS4 (upper panel) or to I5 via TS5
      (lower panel).}
    \label{sifig:I2_traj_2}
\end{figure}

\begin{figure} [H]
    \centering \includegraphics[width=0.8\linewidth]{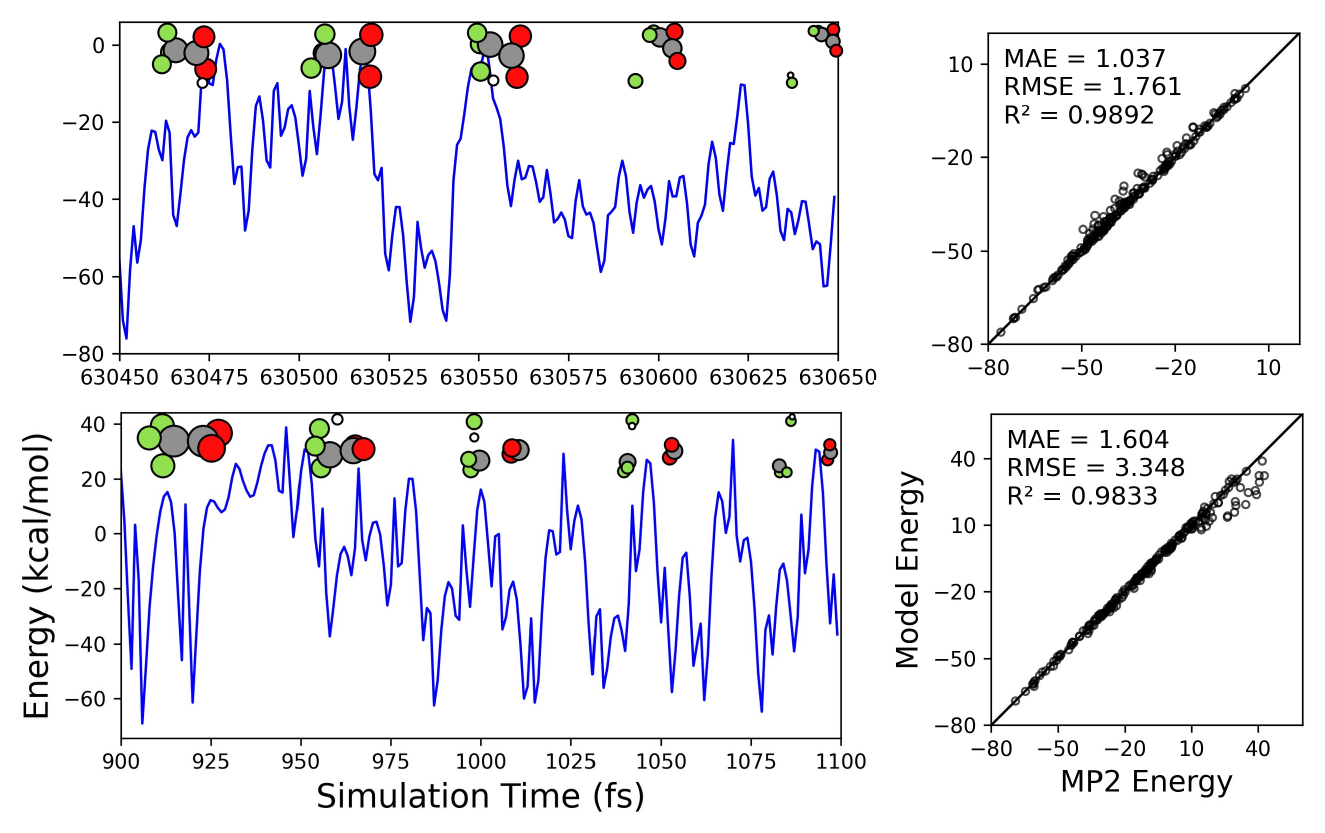}
    \caption{Two representative trajectories following I4 are shown. In the first,
      the system proceeds through TS7 to form HF+CO$_2$CF$_2$ (upper
      panel). With the excess energy available, CO$_2$CF$_2$ can
      further decompose into CO$_2$+CF$_2$ (lower panel).}
    \label{sifig:I4_traj}
\end{figure}

\begin{figure} [H]
    \centering \includegraphics[width=0.8\linewidth]{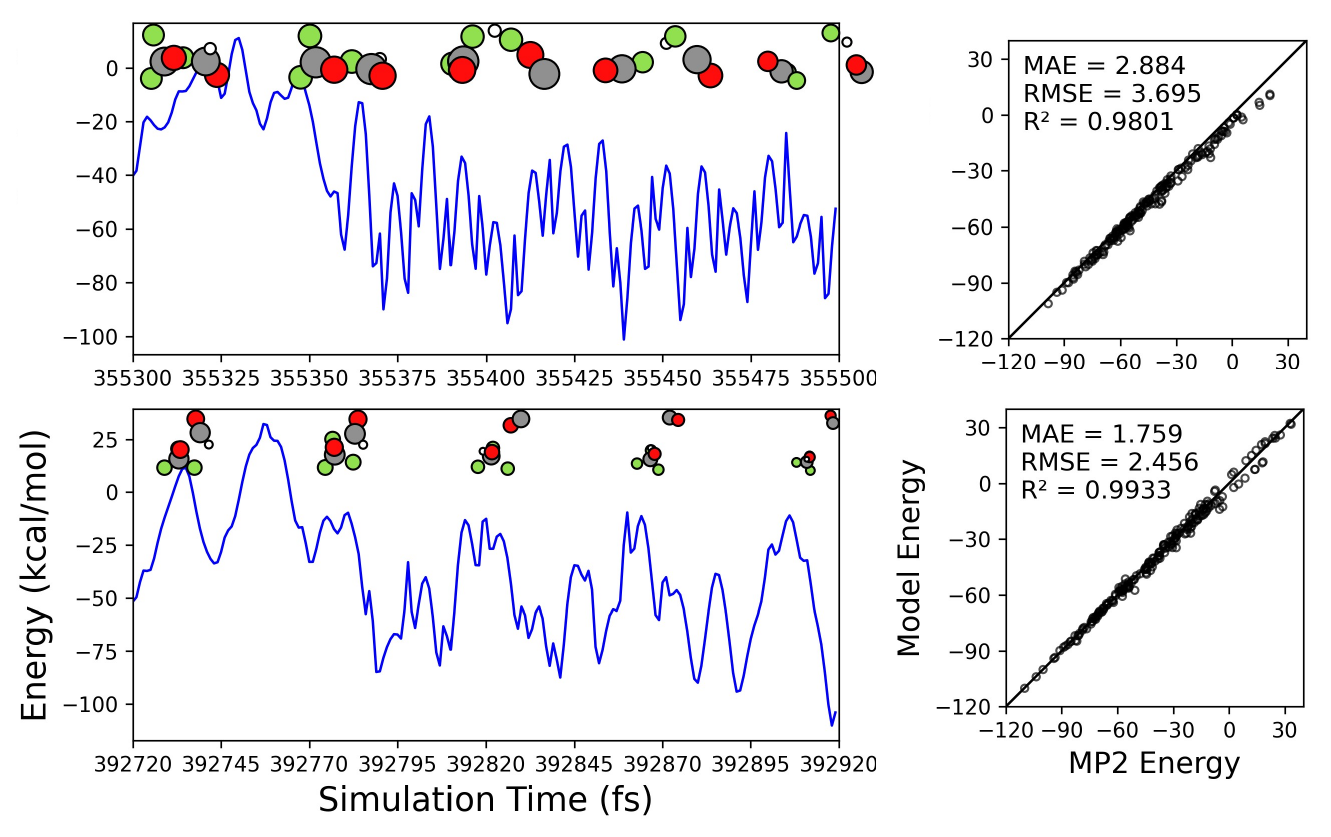}
    \caption{Two representative trajectories show that I5 can
      decompose either into HF+CO+CF$_2$O via TS8 (upper panel) or
      into CO+CF$_3$OH via TS9 (lower panel).}
    \label{sifig:I5_traj}
\end{figure}

\begin{figure}[H]
    \centering
    \includegraphics[width=0.8\linewidth]{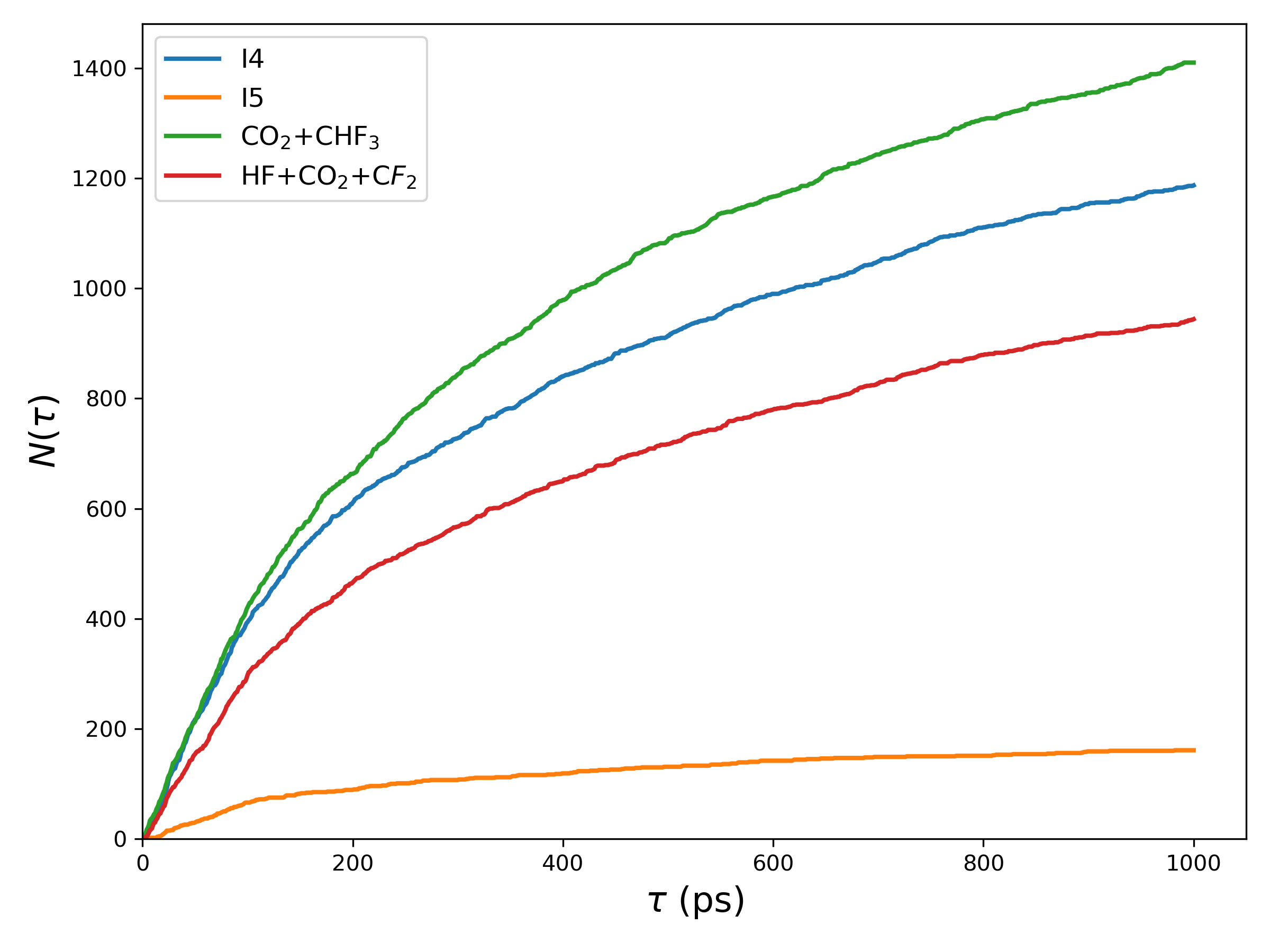}
    \caption{The cumulative populations of the intermediates I4 and
      I5, and the major products CO$_2$+CHF$_3$ and HF+CO$_2$+CF$_2$,
      as functions of time $\tau$, obtained from 10000 molecular
      dynamics trajectories.}
    \label{sifig:population_vs_tau}
\end{figure}

\begin{figure} [H]
    \centering \includegraphics[width=0.8\linewidth]{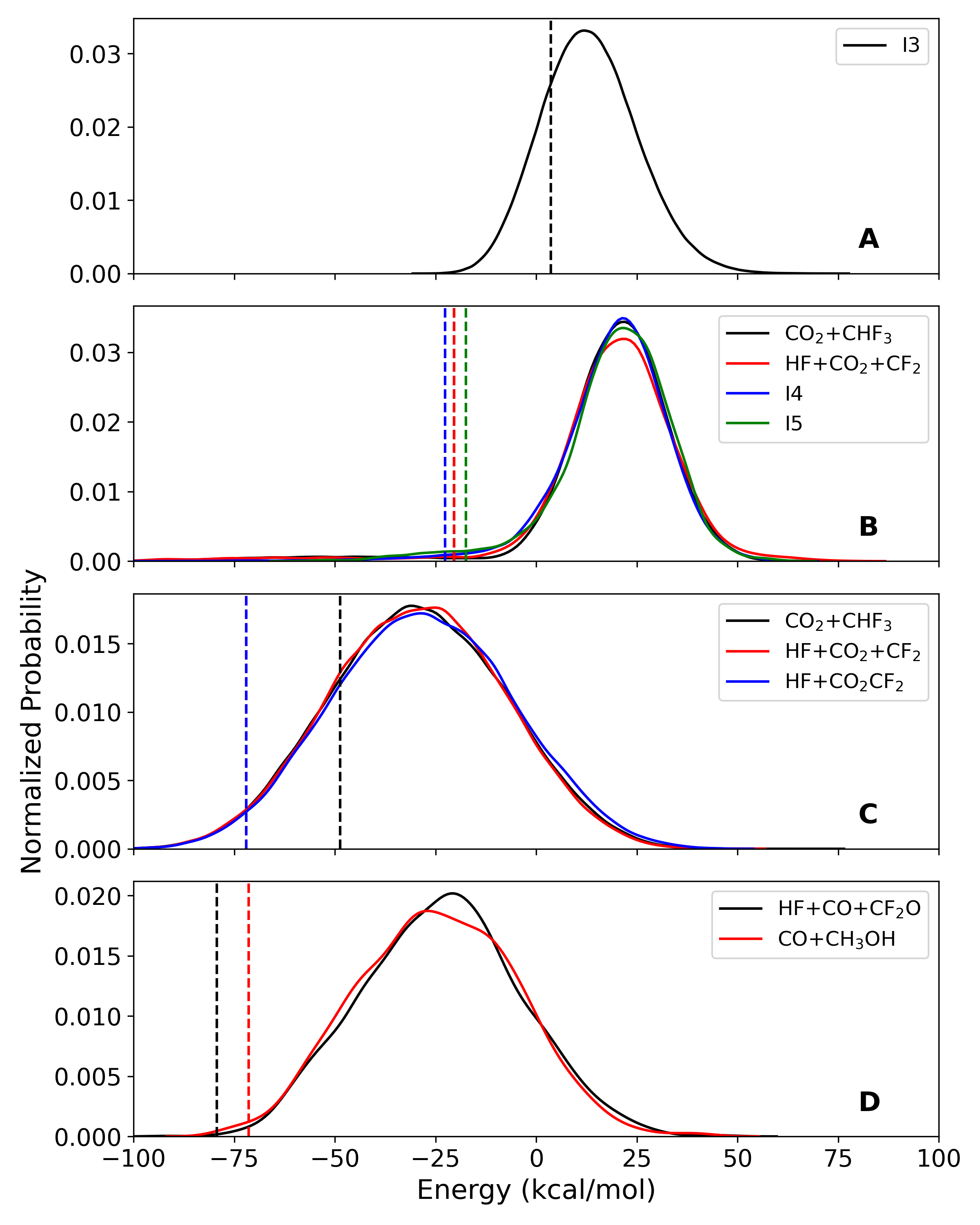}
    \caption{Potential energy distributions of intermediates I2 (panel
      A), I3 (panel B), I4 (panel C), and I5 (panel D), corresponding
      to the reaction channels shown in Figure
      \ref{fig:reaction_diagram}. Different colors denote distinct
      exit channels of each intermediate. The TS energies are
      indicated by dashed vertical lines in the corresponding color,
      respectively.}
    \label{sifig:energy}
\end{figure}

\begin{figure} [H]
    \centering \includegraphics[width=0.8\linewidth]{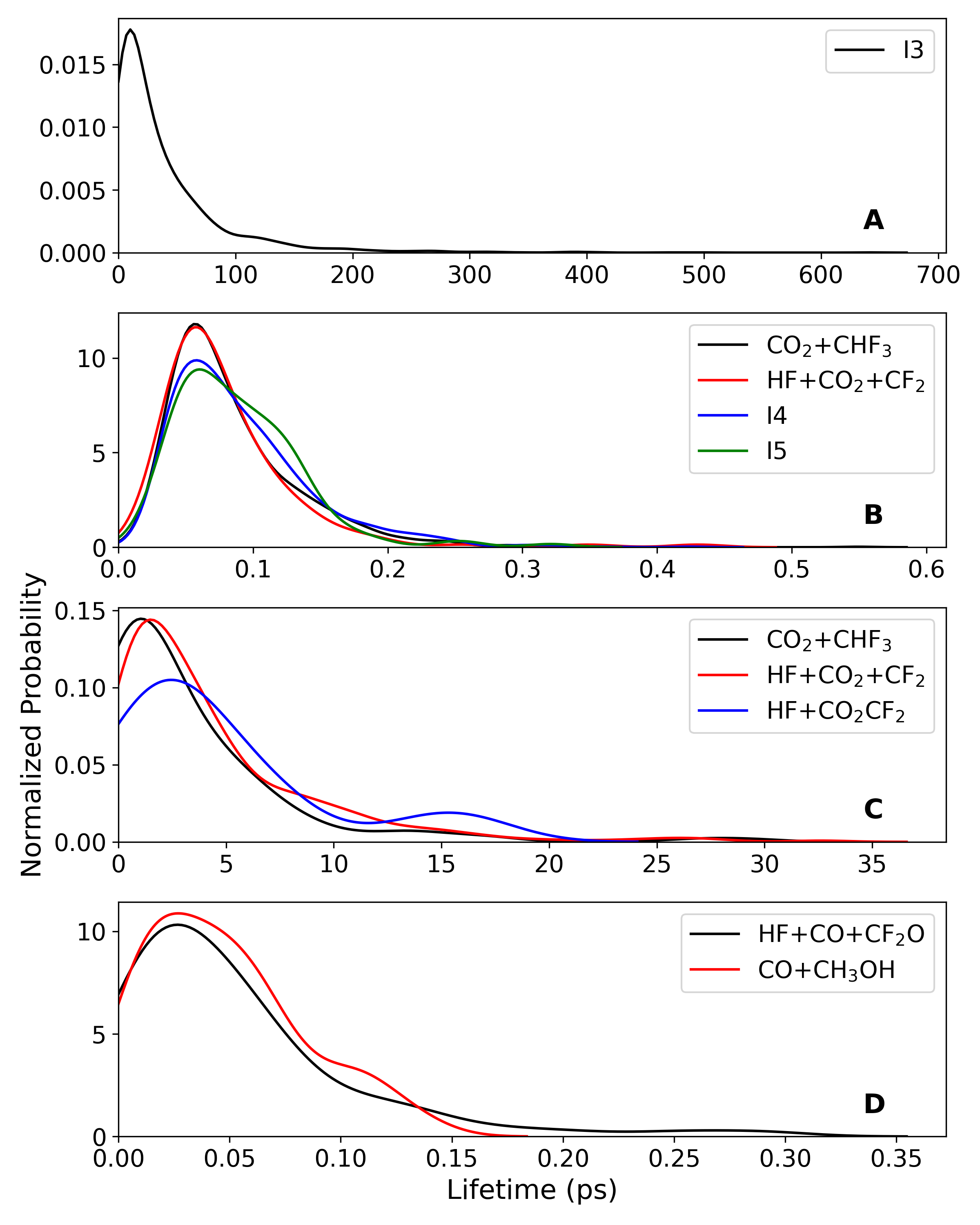}
    \caption{Lifetime distributions of intermediates I2 (panel A), I3
      (panel B), I4 (panel C), and I5 (panel D), corresponding to the
      reaction channels shown in Figure
      \ref{fig:reaction_diagram}. Different colors denote distinct
      exit channels of each intermediate.}
    \label{sifig:life_kde}
\end{figure}

\begin{figure} [H]
    \centering \includegraphics[width=0.8\linewidth]{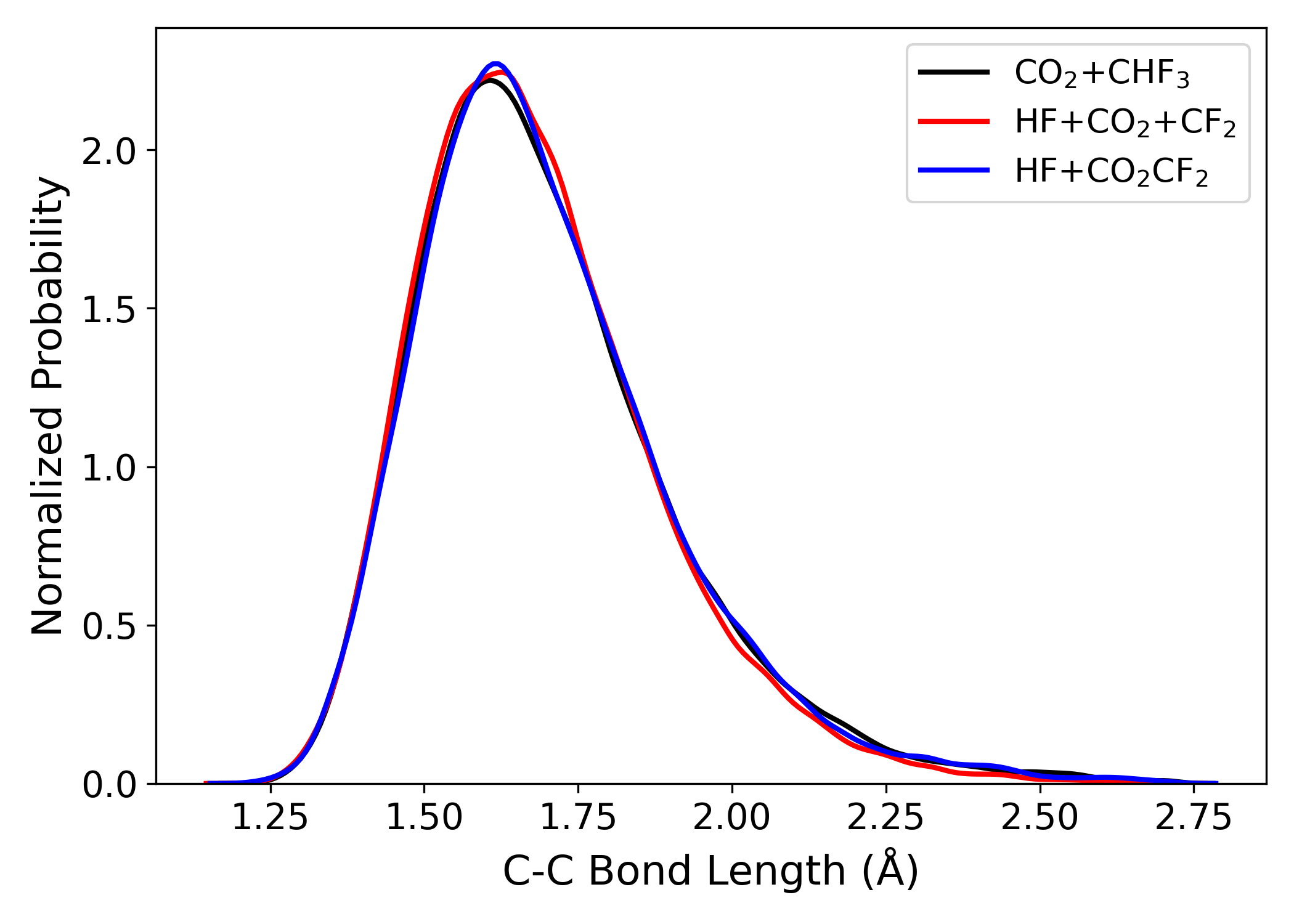}
    \caption{Normalized probability distributions $P(r_{\rm CC})$ sampled in the I4 well prior to accessing the three product channels shown in Figures \ref{fig:reaction_diagram}. The distributions for all three cases are essentially indistinguishable, indicating no significant channel-dependent bias in the C–C bond length distribution at this stage.}
    \label{sifig:I4_coor}
\end{figure}

\section{Tables}

\begin{table} [H]
    \centering
    \begin{tabular}{c|c|c|c|c|c}
    \hline
    Evaluation & ${\rm MAE}(E)$ & ${\rm RMSE}(E)$ & ${\rm MAE}(F)$ & ${\rm RMSE}(F)$ & $R^2 (E)$\\
    \hline
    Model1 & 0.23 & 1.10 & 0.31 & 1.29 & 0.99967\\
    Model2 & 0.32 & 0.72 & 0.36 & 1.90 & 0.99986\\
    Model3 & 0.26 & 0.81 & 0.29 & 0.63 & 0.99982\\
    Model4 & 0.20 & 0.53 & 0.30 & 1.68 & 0.99993\\
    \hline
    \end{tabular}
    \caption{Test set performance of the trained PhysNet models
      evaluated on approximately 2000 structures. Reported metrics
      include mean absolute error (MAE) and root-mean-square error
      (RMSE) for both energies and forces, as well as the coefficient
      of determination ($R^2$) for energies. Energies are given in
      kcal/mol and forces in kcal/mol/\AA.}
    \label{sitab:evaluation}
\end{table}

\begin{table} [H]
    \centering
    \begin{tabular}{lclc}
\hline
Product & Yield (\%) & Product & Yield (\%) \\
\hline
Criegee & 70.4 $\pm$ 0.6 & CO$_2$+CHF$_3$ & 14.1 $\pm$ 0.5 \\
HF+CO$_2$+CF$_2$ & 9.4 $\pm$ 0.4 & HF+CO+CF$_2$O & 0.9 $\pm$ 0.2 \\
\textcolor{blue}{COOH+CF$_3$} & 0.8 $\pm$ 0.1 & \textcolor{blue}{CF$_2$+FCOOH} & 0.8 $\pm$ 0.1 \\
\textcolor{blue}{OH+CO+CF$_3$} & 0.7 $\pm$ 0.1 & dioxi & 0.6 $\pm$ 0.1 \\
CO+CF$_3$OH & 0.5 $\pm$ 0.1 & \textcolor{blue}{H+CO$_2$+CF$_3$} & 0.5 $\pm$ 0.1 \\
HF+CO$_2$CF$_2$ & 0.3 $\pm$ 0.1 & \textcolor{blue}{OH+CF$_3$CO} & 0.3 $\pm$ 0.1 \\
other & 0.6 $\pm$ 0.1 & explosion & 0.1 $\pm$ 0.0 \\
\hline
    \end{tabular}
    \caption{Fractional product yields from 10000 MD
      simulations. Initial structures were sampled at 300K;
      Temperature: 16000K; Timestep: 0.1 fs; Simulation time: 1
      ns. The product CO$_2$+CHF$_3$ is formed directly through I3
      (12.6 \%), and through I4 (1.5 \%). The product HF+CO$_2$+CF$_2$
      is generated from dynamics through I3 (1.5 \%) and through I4
      (7.9 \%). The bootstrapping was performed to obtain the
      uncertainties. 5000 trajectories were randomly sampled with
      replacement 1000 times. The mean fraction and standard deviation
      were computed from these bootstrap samples. The small
      uncertainties indicate reliable convergence. All products are
      listed in the table, including those not shown in the reaction
      diagram (highlighted in blue). These species are high-energy
      radicals, and the MP2 method is known to provide an inadequate
      description.}
    \label{sitab:yield}
\end{table}

\begin{table} [H]
    \centering
    \begin{tabular}{c|ccccccccc}
    \hline
    Species & I1 & I2 & I3 & I4 & I5 & & & & \\
    T1 value & 0.029 & 0.015 & 0.021 & 0.016 & 0.015 & & & & \\
    \hline
    Species & TS1 & TS2 & TS3 & TS4 & TS5 & TS6 & TS7 & TS8 & TS9 \\
    T1 value & 0.040 & 0.025 & 0.020 & 0.022 & 0.023 & 0.018 & 0.021 & 0.018 & 0.018 \\
    \hline
    \end{tabular}
    \caption{T1 values for all the minima and TSs studied in this
      work.}
    \label{sitab:T1}
\end{table}

\begin{table} [H]
    \centering
    \begin{tabular}{c|c|c|c|c}
    \hline
     & VDZ(10e9o) & VDZ(8e8o) & aVDZ(6e6o) & aVTZ(6e6o) \\
    \hline
    Barrier & 22.22 & 23.84 & 22.51 & 22.48 \\
    \hline
    \end{tabular}
    \caption{TS1 barrier height calculated with CASPT2 at different
      basis set and active space.}
    \label{sitab:caspt2}
\end{table}

\begin{table} [H]
    \centering
    \begin{tabular}{c|c|c|c|c|c|c|c|c|c}
    \hline
     TS & TS1 & TS2$^*$ & TS3 & TS4 & TS5 & TS6 & TS7 & TS8 & TS9 \\
    \hline
    Model4 & 780 & 266 & 791 & 447 & 288 & 2091 & 263 & 854 & 1188 \\
    MP2 & 778 & 210 & 794 & 465 & 293 & 2093 & 267 & 833 & 1180 \\
    \hline
    \end{tabular}
    \caption{The imaginary frequencies for each TS, in cm$^{-1}$. For
      TS2, the asterisk indicates that it was not possible to
      determine this TS structure from conventional {\it ab initio}
      calculations, such as QST2 or QST3. Rather, the structure was
      obtained from the trained ML-PES and the frequency was computed
      from MP2 calculations on this structure.}
    \label{sitab:freq}
\end{table}

\begin{longtable}{c c l r}
\caption{Product channels and relative MP2 energies. The
  yellow-highlighted channels are those analyzed in detail, while the
  blue-highlighted channels correspond to radical products listed in
  the Table \ref{sitab:yield}. The red-highlighted channels correspond
  to thermodynamically favorable (negative energy) pathways, but they
  were not observed in the MD simulations. Specifically, for channel
  37, once HF is formed, channels 36 and 38 become energetically more
  accessible than 37; For 52, since three F atoms are initially bonded
  to a single C atom, forming CHF$_2$ requires one F atom to be
  replaced by a hydrogen from another C atom. This exchange is
  geometrically inaccessible; For 53, similar to 52, but the exchange
  occurs between F and O atoms instead of F and H atoms.}\\

\hline
Channel ID & $n_{\rm products}$ & Products & Relative Energy (kcal/mol) \\
\hline
\endfirsthead

\hline
Channel ID & $n_{\rm products}$ & Products & Relative Energy (kcal/mol) \\
\hline
\endhead

1 & 3 & O$_2$ + F$_2$ + C$_2$HF & 160.77 \\
2 & 3 & O$_2$ + HF + C$_2$F$_2$ & 65.43 \\
3 & 3 & O$_2$ + CF + CHF$_2$ & 144.80 \\
4 & 3 & O$_2$ + CH + CF$_3$ & 168.94 \\
5 & 3 & O$_2$ + CF$_2$ + CHF & 140.33 \\
6 & 2 & O$_2$ + C$_2$HF$_3$ & 10.33 \\
7 & 3 & OF + OF + C$_2$HF & 226.64 \\
8 & 3 & OF + F$_2$ + C$_2$HO & 220.69 \\
9 & 3 & OF + OH + C$_2$F$_2$ & 179.67 \\
10 & 3 & OF + HF + C$_2$FO & 130.16 \\
11 & 3 & OF + CO + CHF$_2$ & 89.40 \\
12 & 3 & OF + CF + CHFO & 135.66 \\
13 & 3 & OF + CH + CF$_2$O & 168.01 \\
14 & 3 & OF + CFO + CHF & 176.13 \\
15 & 3 & OF + CF$_2$ + CHO & 134.41 \\
16 & 3 & OF + OF$_2$ + C$_2$H & 328.40 \\
17 & 2 & OF + C$_2$HF$_2$O & 83.63 \\
18 & 3 & F$_2$ + OH + C$_2$FO & 178.54 \\
19 & 3 & F$_2$ + HF + C$_2$O$_2$ & 84.22 \\
20 & 3 & F$_2$ + CO + CHFO & 14.39 \\
21 & 3 & F$_2$ + CF + CHO$_2$ & 154.99 \\
22 & 3 & F$_2$ + CH + CFO$_2$ & 206.04 \\
23 & 3 & F$_2$ + FO$_2$ + C$_2$H & 301.88 \\
24 & 3 & F$_2$ + HO$_2$ + C$_2$F & 261.88 \\
25 & 3 & F$_2$ + CO$_2$ + CHF & 87.32 \\
26 & 3 & F$_2$ + CFO + CHO & 104.34 \\
27 & 2 & F$_2$ + C$_2$HFO$_2$ & 18.56 \\
\rowcolor{blue}
28 & 3 & OH + CO + CF$_3$ & 4.03 \\
29 & 3 & OH + CF + CF$_2$O & 58.48 \\
30 & 3 & OH + CFO + CF$_2$ & 57.99 \\
31 & 3 & OH + OF$_2$ + C$_2$F & 271.98 \\
\rowcolor{blue}
32 & 2 & OH + C$_2$F$_3$O & -4.35 \\
\rowcolor{yellow}
33 & 3 & HF + CO + CF$_2$O & -111.15 \\
34 & 3 & HF + CF + CFO$_2$ & 48.15 \\
35 & 3 & HF + FO$_2$ + C$_2$F & 197.09 \\
\rowcolor{yellow}
36 & 3 & HF + CO$_2$ + CF$_2$ & -79.20 \\
\rowcolor{red}
37 & 3 & HF + CFO + CFO & -20.45 \\
\rowcolor{yellow}
38 & 2 & HF + CO$_2$CF$_2$ & -81.77 \\
39 & 3 & CO + OF$_2$ + CHF & 168.30 \\
40 & 3 & CO + HOF + CF$_2$ & 42.17 \\
\rowcolor{yellow}
41 & 2 & CO + CF$_3$OH & -119.41 \\
42 & 3 & CF + FO$_2$ + CHF & 263.04 \\
43 & 3 & CF + HO$_2$ + CF$_2$ & 161.31 \\
44 & 3 & CF + CHO + OF$_2$ & 217.77 \\
45 & 3 & CH + FO$_2$ + CF$_2$ & 254.42 \\
46 & 3 & CH + OF$_2$ + CFO & 250.87 \\
\rowcolor{blue}
47 & 2 & CF$_2$ + FCOOH & -63.97 \\
48 & 2 & FO$_2$ + C$_2$HF$_2$ & 134.63 \\
49 & 2 & HO$_2$ + C$_2$F$_3$ & 108.73 \\
\rowcolor{yellow}
50 & 2 & CO$_2$ + CHF$_3$ & -140.90 \\
51 & 2 & CFO + CF$_2$OH & 6.02 \\
\rowcolor{red}
52 & 2 & CHF$_2$ + CFO$_2$ & -0.55 \\
\rowcolor{red}
53 & 2 & CF$_2$O + CHFO & -113.60 \\
\rowcolor{blue}
54 & 2 & COOH + CF$_3$ & -27.46 \\
\hline
\label{sitab:channels}
\end{longtable}

\bibliography{refs.clean}

\end{document}